# From Impurity Doping to Metallic Growth in Diffusion Doping: Properties and Structure of Ag Doped InAs Nanocrystals

*Yorai Amit[1,2], Yuanyuan Li[3], Anatoly I. Frenkel[*,3] and Uri Banin[*,1,2]*

1 The Institute of Chemistry, Hebrew University, Jerusalem 91904, Israel.

2 The Center for Nanoscience and Nanotechnology, Hebrew University, Jerusalem 91904, Israel.

3 Department of Physics, Yeshiva University, New York, New York 10016, United States

* Address correspondence to: anatoly.frenkel@yu.edu, uri.banin@mail.huji.ac.il

**TOC Figure**

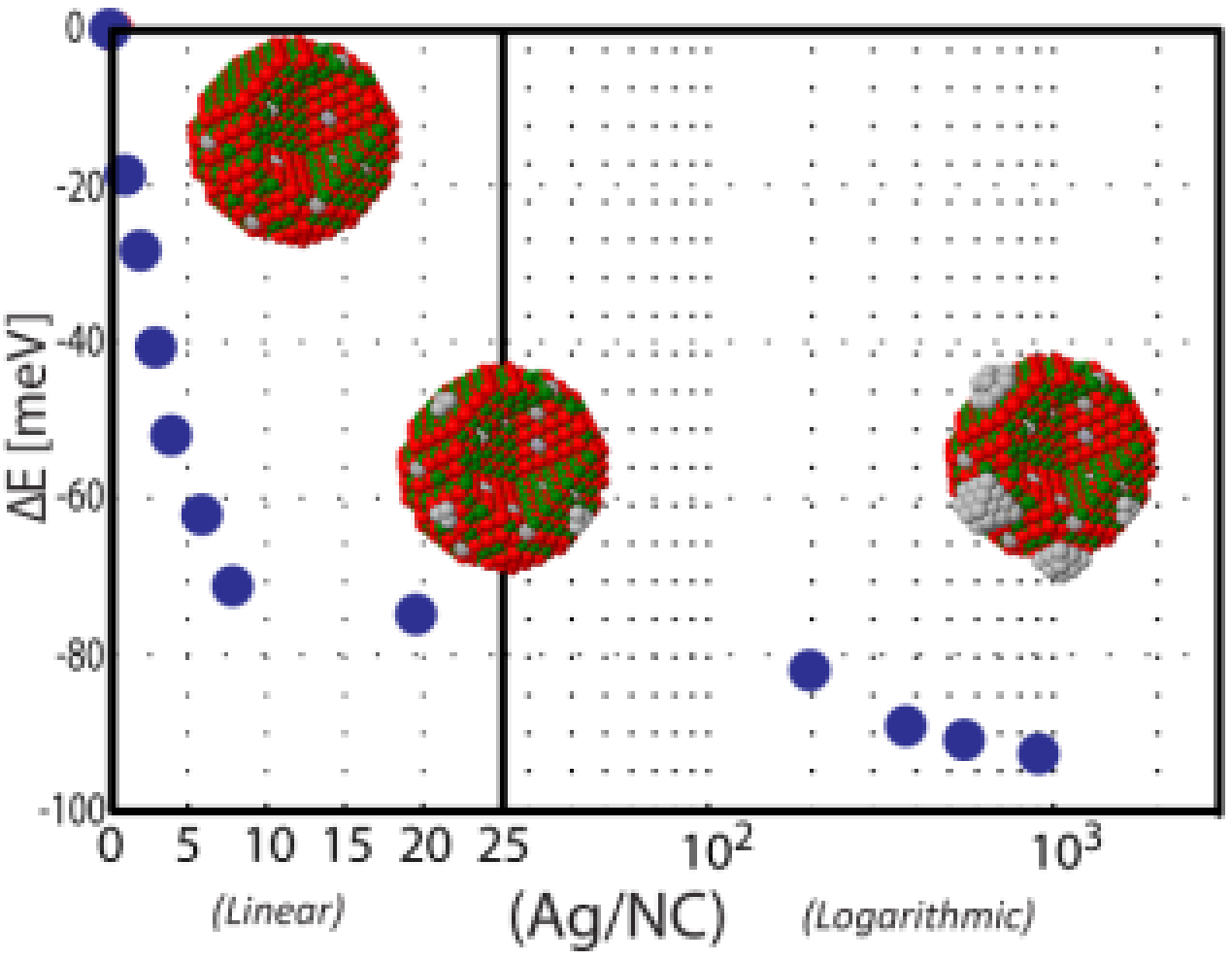

**Abstract.** Tuning of the electronic properties of pre-synthesized colloidal semiconductor nanocrystals (NCs) by doping plays a key role in the prospect of implementing them in printed electronics devices such as transistors, and photodetectors. While such impurity doping reactions have already been introduced, the understanding of the doping process, the nature of interaction between the impurity and host atoms, and the conditions affecting the solubility limit of impurities in nanocrystals are still unclear. Here, we used a post-synthesis diffusion based doping reaction to introduce Ag impurities into InAs NCs. Optical absorption spectroscopy along with analytical inductively coupled plasma mass-spectroscopy (ICP-MS) were used to present a two stage doping model consisting of a "doping region" and a "growth region", depending on the concentration of the impurities in the reaction vessel. X-ray absorption fine-structure (XAFS) spectroscopy was employed to determine the impurity location and correlate between the structural and electronic properties for different sizes of InAs NCs and dopant concentrations. The resulting structural model describes a heterogeneous system where the impurities initially dope the NC, by substituting for In atoms near the surface of the NC, until the "solubility limit" is reached, after which the rapid growth and formation of metallic structures are identified.

Colloidal semiconductor nanocrystals (NCs) manifest unique size dependent optoelectronic properties.[1,2] Advancements in the synthesis of semiconductor NCs enable the use of different semiconductor materials,[3-5] the fabrication of highly controlled shapes,[6-8] and various compositions,[9-11] as means of further controlling their electronic properties. The ability to tune the electronic properties of semiconductor NCs has led to the demonstrations of various applications based on NC building-blocks.[12-14] The process of doping, which is well known for bulk semiconductors, presents an additional method for controlling the physical properties of semiconductor materials and as a result led to the demonstration of numerous optoelectronic applications.[15] Therefore, great efforts have been put into the prospect of doping semiconductor colloidal nanocrystals.[16-20] To this end, the doping of semiconductor NCs has been demonstrated *via in-situ* methods, by adding impurities to the reaction vessel[21-26] or *ex-situ* such as electro-chemical reactions,[27,28] thermal treatments,[29,30] and the more conventional impurity doping process, where the impurities are introduced into the semiconductor lattice post-synthesis.[31-33] However, impurity doping of nanocrystals is still considered a difficult task, mainly due to "self-purification" processes, owing to the small size of the NC, resulting in the expulsion of the impurities to the surface of the NC.[34-36] This was previously described for InAs NCs, where the root-mean-square distance that a diffusing atom would travel in 24 hours was found to greatly exceed the size of the NCs.[31] This difficulty was also shown for P doping of Si NCs.[37] Here, a theoretical study revealed a critical size of the NC host, below which the P atoms are expelled to the surface of the NCs.

We have previously presented the synthesis of heavily doped InAs colloidal NCs by means of a room-temperature, solution-phase, diffusion based reaction, in which Ag and Cu impurities were introduced to a solution of pre-synthesized NCs.[31] Scanning-tunneling-spectroscopy (STS) measurements of Cu (Ag) doped NCs revealed a blue (red) shift of the Fermi energy proving the *n*-type (*p*-type) doping. It was also shown that neither the crystal structure nor the size of the NCs was affected by the doping reaction, ruling out the possibility of size-dependent effects governing the observed changes in the optoelectronic properties. In a following work, we have reported on the detailed characterization of Cu doping of InAs NCs, using advanced x-ray absorption fine structure (XAFS) spectroscopy to establish the location of the impurity and its electronic state within the NC as a function of the impurity concentration in the reaction solution.[38] We found that the diffusion of Cu impurities into the NC lattice is favored and results in an energetically stable system. Furthermore, we identified the impurity site determining that the Cu impurities occupy only hexagonal-interstitial sites throughout the lattice for a very wide range of impurity concentrations. This is fully consistent with the *n*-type behavior measured for the Cu doped InAs both using STM and in optical measurements.[31]

Recently, Norris *et al.* reported on the Ag doping of CdSe NCs, using also a post-synthesis solution-phase diffusion doping reaction.[32] In their work they have found that the electronic properties of the doped CdSe NCs shift from an *n*-type to a *p*-type behavior, as a result of the impurity assuming initially an interstitial site followed by its transformation into a substitutional lattice site, depending on the impurity concentration in the lattice. Subsequent modeling of the system helped to better understand the doping mechanism and the suggested structural transformations. The modeling results indicate that the Ag impurities are initially able to penetrate the lattice by diffusion through interstitial sites, followed by their repulsion to the surface of the NC where they assume substitutional sites, through the kick-out of the Cd host atoms, and eventually forming impurity-pairs in both substitutional and interstitial lattice sites.[39]

While Ag exhibits a bimodal behavior in CdSe, the optoelectronic studies of Ag doping in InAs suggests a purely *p*-type behavior.[31] Further structural analysis is needed in order to fully understand the doping mechanism and the nature of interactions between the impurity and the host in such a system. In our previous work on Cu doped InAs NCs we addressed some of the intriguing questions dealing with impurity doping in colloidal NCs. First, we have shown that the Cu impurity atoms saturate the lattice by occupying nearly all-possible interstitial sites, keeping the host lattice intact, regardless of the Cu concentration. In light of this, it is interesting to realize how the use of a substitutional impurity, such as Ag, would affect the InAs NC lattice. A second aspect is that of the solubility limit. Cu impurities exhibited no solubility limit in InAs, up to the point of nearly saturating the NC lattice. Ag impurity studied here is much larger than Cu, may have different valence and thus may reveal a different behavior in terms of its solubility. Furthermore, in NCs, the solubility limit is of great interest due to the possible size effects.

In the current work we present a size-dependent study of Ag doped InAs NCs using both analytical and spectroscopic methods to better understand the doping mechanism and correlate between the observed changes in the optoelectronic properties and the local environment around the impurity atom for different impurity concentrations. Furthermore, using analytical inductively coupled plasma mass-spectrometry (ICP-MS) measurements together with optical absorption measurements, we were able to identify two regions: a "doping region" and a "growth region" that depend strongly on the concentration of impurities in the reaction solution, and points towards an identification of a "solubility limit" of Ag impurities in InAs NCs.

## Results and Discussion:

**Identification of the impurity “solubility limit”.** Colloidal InAs NCs were synthesized following a well-established synthesis[40] followed by a size-selective precipitation to obtain narrow size distributions. The pre-synthesized InAs NCs were then doped with various amounts of Ag impurities by adding calculated volume fractions of the impurity solution containing the metal salt ($AgNO_3$), Didodecyldimethylammonium bromide (DDAB, to stabilize the metal salt in the organic solution) and Dodecylamine (DDA, which stabilizes the NCs and acts as a reducing agent). Figure S1 of the supporting information presents a set of doping solutions, using the different ligands to stabilize the metal salt, (a) at the time of preparation and (b) after 16 hours. TEM image, of the resulting product of the $AgNO_3$@DDA solution after 16 hours (Fig. S2a), and the XRD diffractogram (Fig. S2b) indicate the formation of metallic Ag nanoparticles at room-temperature clearly reflecting the role DDA plays in silver reduction. ,. With DDAB, Ag nanoparticles are not formed indicating its role in stabilizing Ag(I) in solution. For additional information see experimental section.

TEM images of the as-synthesized InAs NCs (Fig. 1a), which are 4.8nm in diameter, and the doped InAs NCs (Fig. 1b), show that neither the size nor the shape of the NCs are changed upon doping, consistent with our earlier study[31], see SI Fig S3 for size histograms and additional TEM images of 4.8nm NCs before and after doping. Figure 1c presents the normalized optical absorption spectra for the 4.8nm InAs NCs upon addition of various ratios of Ag impurities per NC. The data is shown as a function of the molar ratio of Ag to NCs in solution, in the initial stage, denoted as $(Ag/NC)_i$. As can be seen, the position of the first exciton peak gradually red-shifts[41] when increasing the $(Ag/NC)_i$ ratio in the solution. Previous study using STM, optical spectroscopy and theoretical analysis, showed that the doping leads to a *p*-type behavior. Specifically, the red shifted absorption has been interpreted as evidence of band tailing effects related to the disorder induced by impurity insertion into the InAs NCs lattice. For additional information and theoretical modeling of the observed spectral shifts see ref. 31

Figure 1d depicts the shift in the position of the first exciton peak, with respect to the undoped samples, for the 3.4nm (blue), 4.8nm (green), and 6.6nm (red) InAs NCs upon reacting them with various $(Ag/NC)_i$ ratios. (for the optical absorption spectra of the 3.4nm and 6.6nm doped NCs see SI Fig. S4). While all three samples exhibit a red-shift in the electronic structure, differences are noticeable. First, the smaller NCs are more strongly affected by the doping process, completely losing the feature of the first exciton peak when reaching high $(Ag/NC)_i$ ratios (Fig. S4a). The larger NCs, however, do not exhibit the same loss of feature, even when exposed to higher $(Ag/NC)_i$ ratios, although some broadening and loss of structure are observed. It is noted that for the larger 6.6nm NCs (Fig. S4b), the loss of feature of the first exciton peak is induced by ligand absorption at 1200nm (See SI Fig. S5), which intensifies as the doping level is pushed to higher ratios, due to the contribution of the ligands in the impurity solution. The second difference is observed when comparing the extent of the red-shift of the first exciton peak between the three different sizes (Fig. 1d). As the NC grows in size, the electronic shift for a given $(Ag/NC)_i$ ratio is reduced, as well as the overall obtainable shift. For instance, whereas the first exciton peak of the 3.4nm NCs shifts by -70meV for a ratio of 200 $(Ag/NC)_i$, it only reaches values of -35 meV and -16meV for the 4.8nm and 6.6nm NCs, respectively.

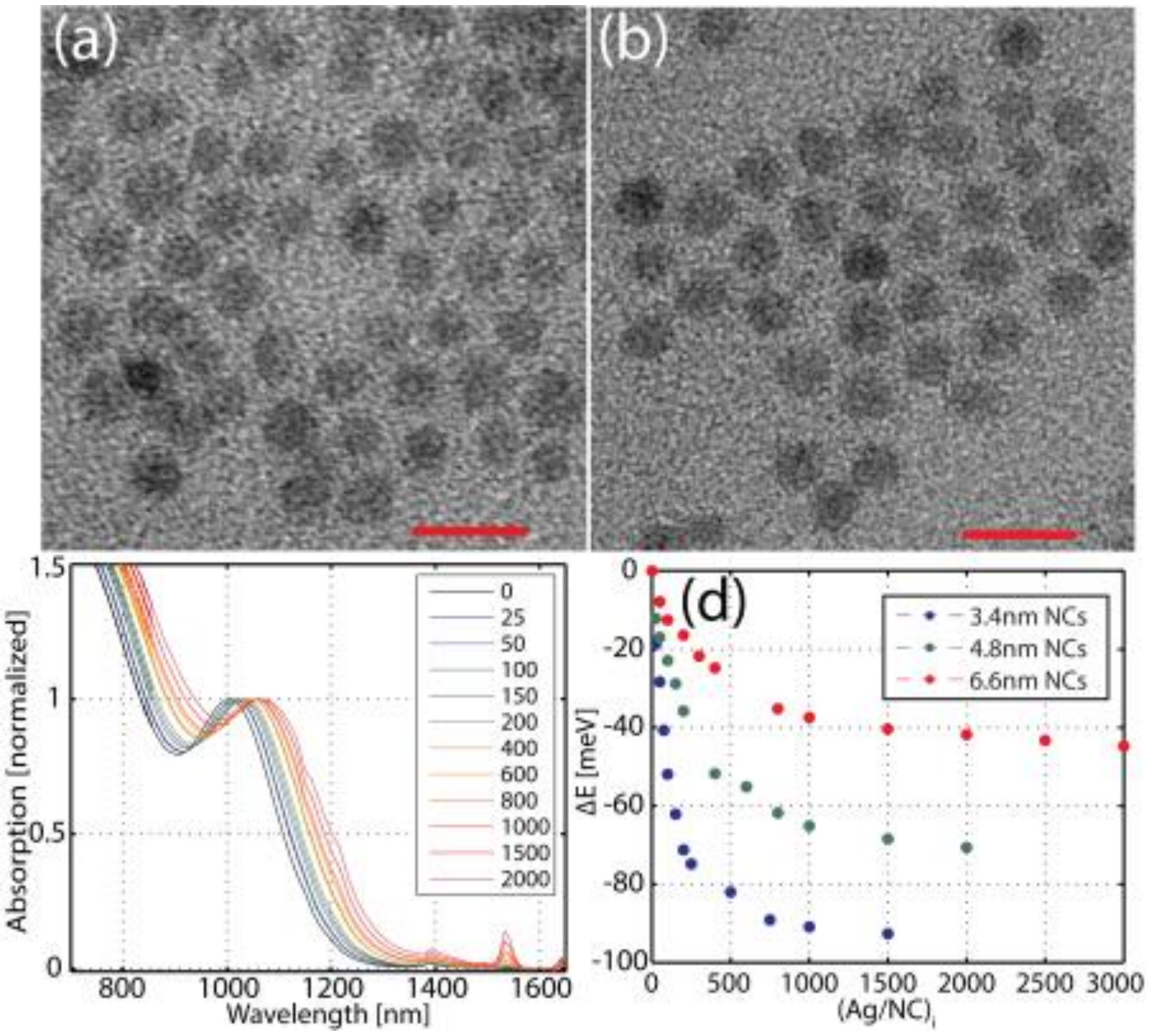

**Figure 1**. TEM images of (a) 4.8nm (diameter) InAs NCs before and (b) after doping to a ratio of 400 $(Ag/NC)_i$ in solution (scale bar = 10nm). (c) Normalized absorption spectra for the 4.8nm NCs for various ratios of up to 2000 $(Ag/NC)_i$, which denote the ratios of Ag to NC in solution phase. (d) Energy shift of the first exciton peak for 3.4nm (blue), 4.8nm (green), and 6.6nm (red) InAs NCs reacted with increasing amounts of Ag in the solution phase. (For the absorption spectra of the 3.4nm and 6.6nm InAs NCs for the different doping concentrations see SI Fig. S2)

Doping experiments were performed over a wide range of molar concentrations and the measured excitonic shifts were found to, essentially, depend only on the $(Ag/NC)_i$ ratio and not on the absolute concentrations (Fig. S6). Hence, we present our results in terms of the $(Ag/NC)_i$ ratio, which is a convenient description of the doping conditions, as will be shown below.

Inductively Coupled Plasma – Mass Spectrometry (ICP-MS) measurements were conducted to correlate between the $(Ag/NC)_i$ ratio in solution and the actual number of Ag impurities per NC, denoted as $(Ag/NC)_a$. For these measurements, the doped-NCs were isolated from the solution by solvent/anti-solvent precipitation and were then dissolved in concentrated nitric acid ($HNO_3$ 70%) and diluted with triply distilled water (for additional information see experimental section). ICP-MS results (Fig. 2) for a series of 3.4nm (blue), 4.8nm (green), and 6.6nm (red) InAs NC doped with various concentrations of Ag atoms reveal two distinct regions. The first region, observed for the low $(Ag/NC)_i$ ratios, indicates that only few impurities are present in each NC (see Fig. 2 c-d and Fig. S7 for zoom-in view on the doping region limits), and is therefore considered and termed herein as the "doping region", as further established below. The second region, reached when crossing a size-dependent threshold ratio of Ag atoms, is accompanied by a large increase in the $(Ag/NC)_a$ ratio, indicating a strong accumulation of Ag onto the NCs after this threshold, and is considered herein as the "growth regime".

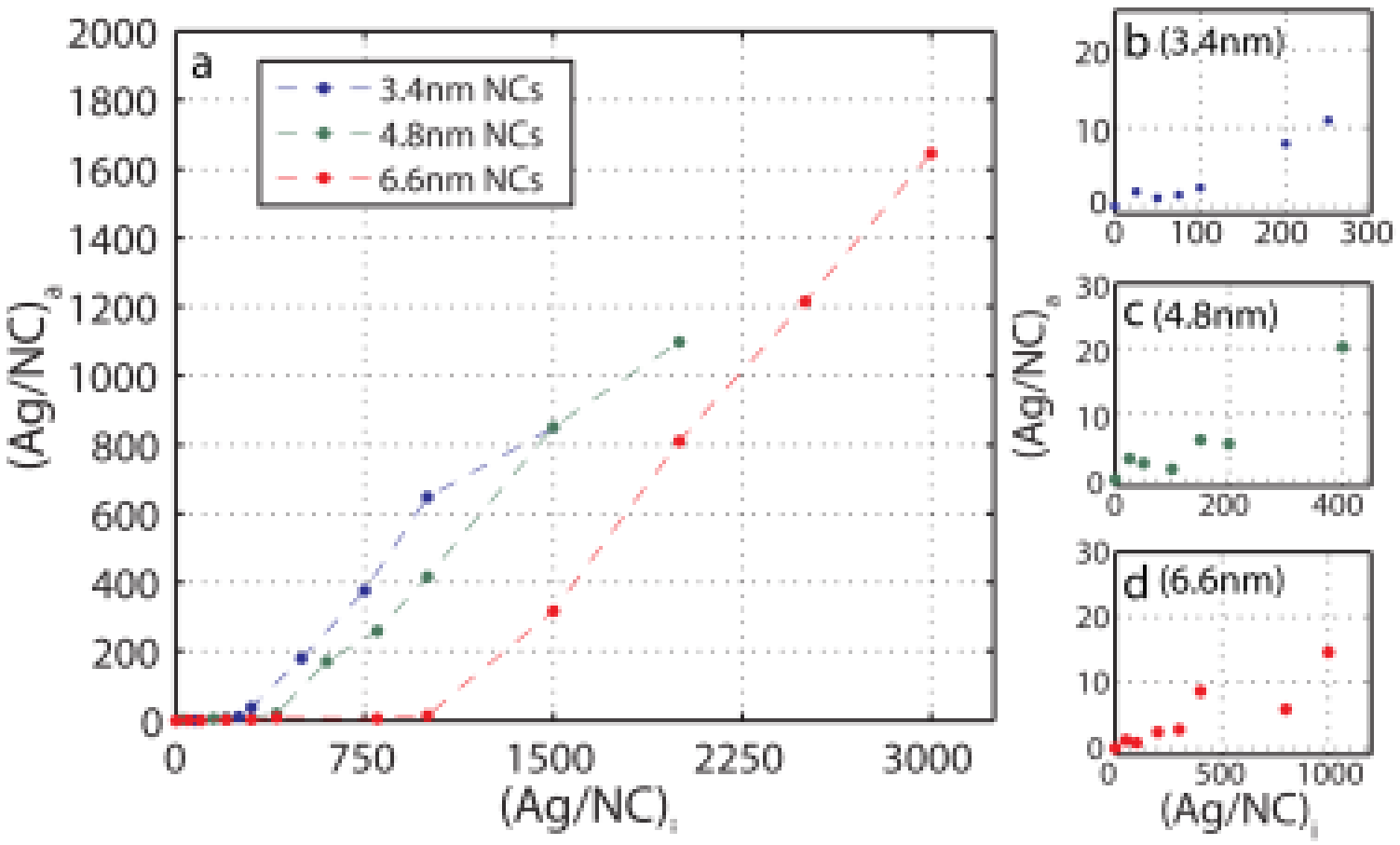

**Figure 2**. Measurement of the $(Ag/NC)_a$ ratio by ICP-MS measurements for 3.4nm (blue), 4.8nm (green), and 6.6nm (red) InAs NCs as a function of the $(Ag/NC)_i$ ratio in solution. (a) shows thew full range of Ag ratios for the three NC sizes, while frames (b), (c) and (d) zoom-in to show the region of low ratios for the three NC sizes.

The onset of the growth phase as a function of the $(Ag/NC)_i$ ratio, was found to increase with NC size (Fig. 2). One question that we can address with the help of the available data is the correlation between the observed crossover value between the two regions and the dimensionality of the NCs. By fitting each portion of the ICP-MS data with a linear function, (Fig. S7) we were able to extract the crossover values for each NC size and found that the transition from the "doping region" to the "growth region" occurs for $(Ag/NC)_i$ ratios of 222±50, 395±55, and 1130±140 for the 3.4nm, 4.8nm, and 6.6nm InAs NCs, respectively. Below these crossover points only few Ag atoms are detected inside the NC (9, 12 or 14 (Fig. S9) as detected by the ICP-MS, while the error bars on these numbers are correspondingly large.

To investigate the relationship between the crossover points (for different particle sizes) and the dimensionality of the NC we plotted their values for each size of NC in terms of the linear, quadratic or cubic functions of the NC radius. A cubic dependence on radius (*i.e.*, volume dependence) provided the best fit (Fig. 3 and SI Fig. S8). Hence, the measured values of the crossover points of the $(Ag/NC)_i$ ratios have a meaning of the solubility limit. Also, its determination through the $(Ag/NC)_i$ ratios in solution as opposed to the $(Ag/NC)_a$ ones determined by ICP-MS is better justified because the former are larger in values and better defined than the latter.

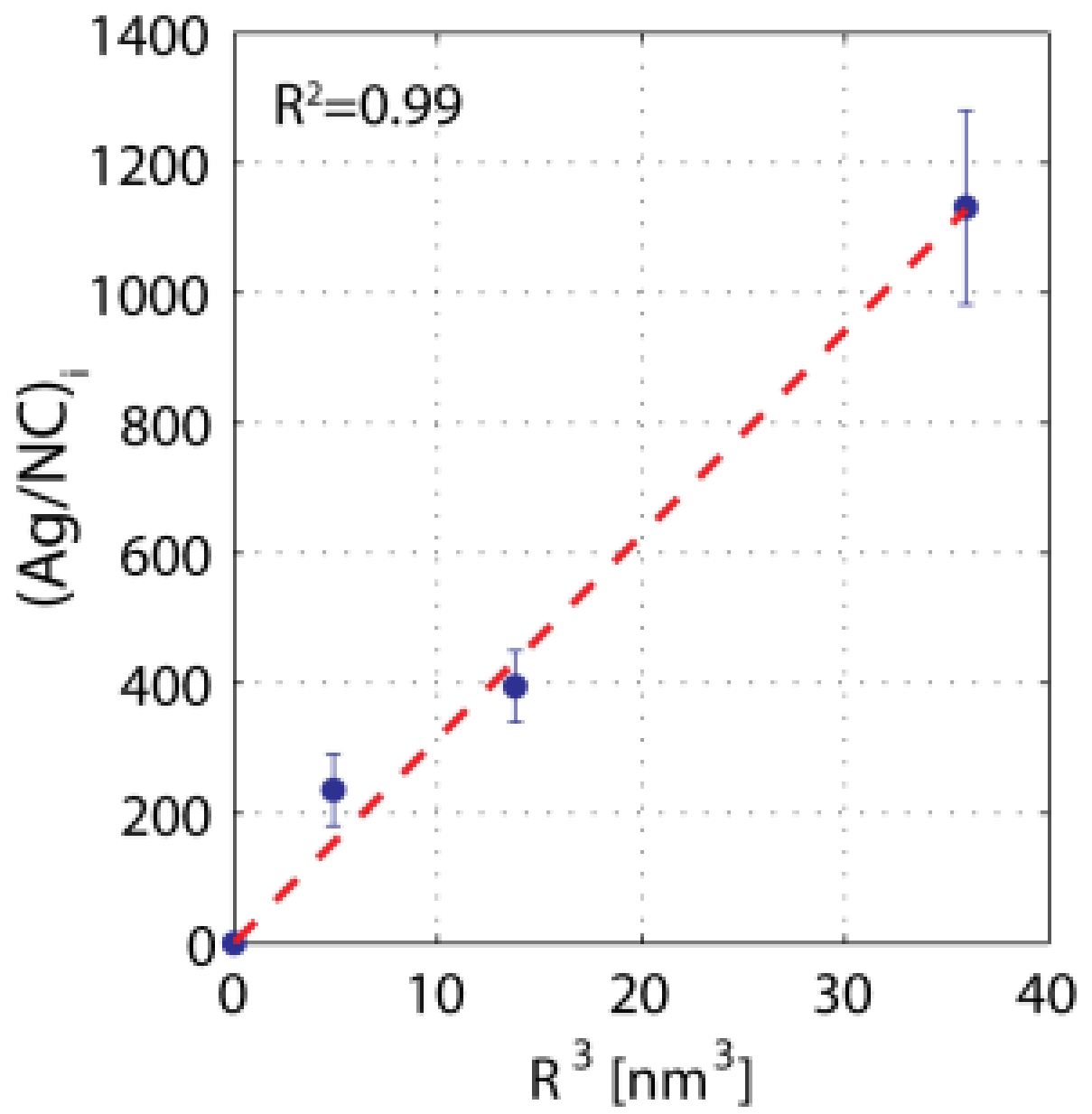

**Figure 3.** Linear fit of the $(Ag/NC)_i$ ratios in the solution, at the onset of the "growth region", as a function of the NC volume for the different NC sizes.

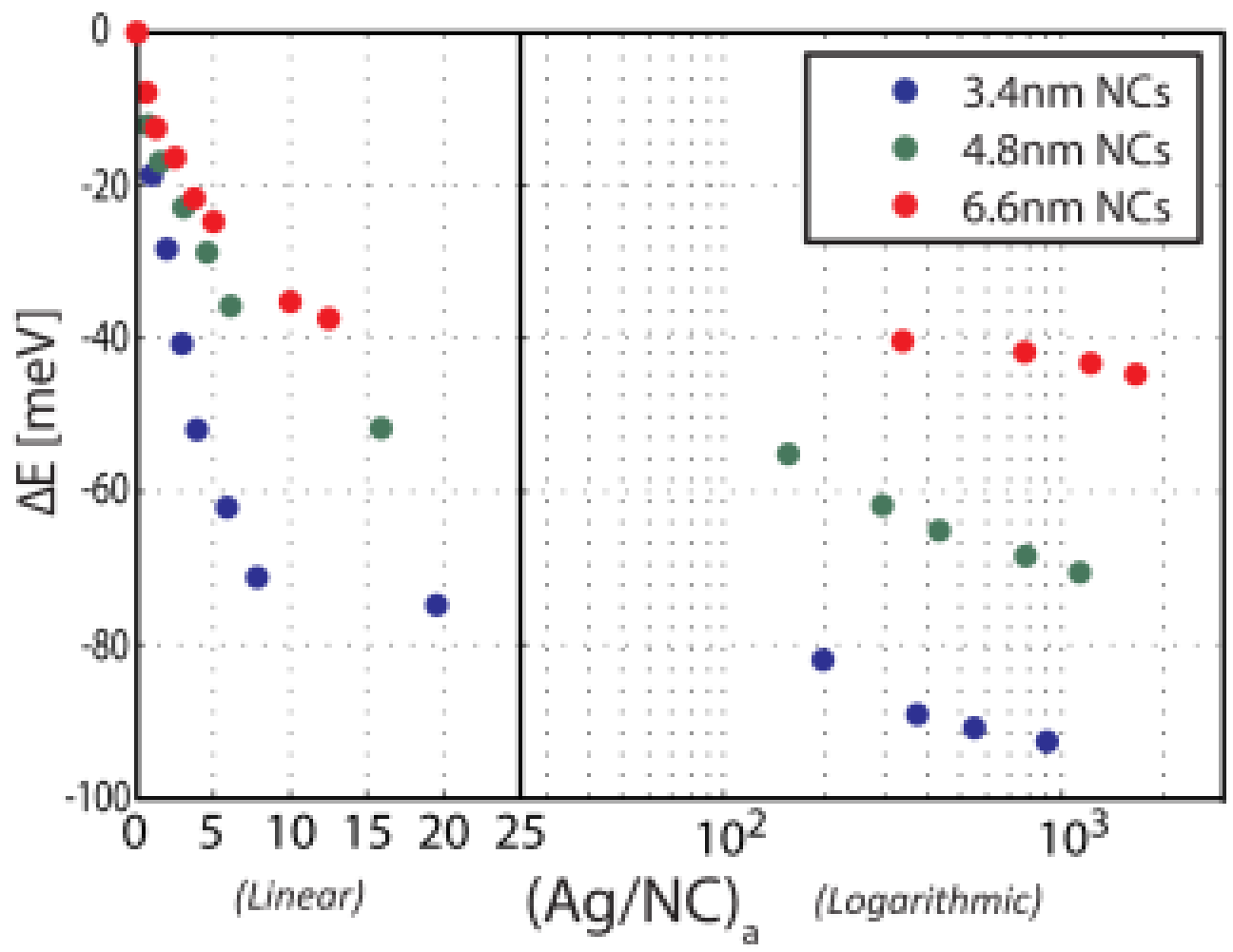

**Figure 4**. Energy shift of the first exciton peak as a function of $(Ag/NC)_a$ for the 3.4nm (blue), 4.8nm (green), and 6.6nm (red) InAs NCs.

Plotting the electronic shift as a function of the $(Ag/NC)_a$ value (Fig. 4) also reveals the two different regions within each size of NCs, similar to the aforementioned ICP-MS data. Here, the first region (*i.e.* the "doping region"), taking place below the estimated "solubility limit", exhibits a substantial effect on the absorption spectrum of the NCs, whereas the second region (*i.e.* the "growth region") has an overall negligible effect on the absorption. For the case of the 3.4nm NCs (blue), the band-gap energy has decreased by approximately 80meV for a doping level of ~8 Ag atoms per NC (equivalent to a concentration of 200 Ag atoms per NC in solution) while it remains relatively unchanged for higher doping levels, decreasing only by an additional 10meV for nearly 1000 Ag atoms per NC. This stands to show that the intersection point of these two regions coincides with the estimated solubility limit for each NC size, as extracted from the ICP-MS measurements.

**XANES and EXAFS data analysis and modeling.** To further investigate the doping mechanism and the formation of two reaction regions indicated by the optical and ICP spectroscopy data, and to elucidate the resulting structure in the doped semiconductor NCs, we employed X-ray absorption fine structure (XAFS) spectroscopy techniques. XAFS methods are sensitive to the local atomic environment and electronic properties of X-ray absorbing atoms and their immediate surroundings, within ca. 0.5-0.8 nm radius from the absorber.[42] They therefore can shed light on the impurity diffusion mechanisms and interactions with the NC host. In our work, XAFS measurements were carried out by depositing the doped NC sample onto a Kapton tape, and mounting the sample into a sealed sample chamber, while in a nitrogen glove-box. The sample chamber was then transferred to the beamline where it was connected to nitrogen flow to maintain inert conditions throughout (see experimental section for more information).

First, in order to better understand the mechanism by which the doping occurs, we have performed solution phase Ag K-edge XAFS measurements on the impurity solution, containing the metal salt and the appropriate ligands (see experimental section for more details), and analyzed the experimental data (Fig. S10). The best fit results for the impurity solution (see SI Table S1) revealed the presence of Ag-Br bonds but no Ag-N, Ag-O, or Ag-Ag bonds, indicating the complete dissociation of the $AgNO_3$ structure in the doping precursor solution and the formation of a ligand-impurity complex with no evidence of metal bonds. The presence of Ag-Br bonds in the doping solution and their absence in the doped samples (see below) sheds light on the mechanism of the doping. As discussed previously, DDAB which acts as a coordinating ligand, stabilizes the Ag(I) species in the organic medium. The Ag impurities adsorb from the solution to the surface of the NC allowing then for its possible entry into the NC. .

To further elucidate the role of each constituent in the doping solution, three solutions of NCs, of similar size and concentration, were doped to the same $(Ag/NC)_i$ using either $AgNO_3$ in Dodecylamine (DDA), $AgNO_3$ in DDAB, or $AgNO_3$ in DDA+DDAB (Fig. S11). The use of an impurity solution containing the metal salt and DDA, resulted in an immediate loss of the first exciton feature with no clear optical absorption shift. Doping with an impurity solution containing only DDAB and the metal salt yields similar optical shifts as those obtained by doping using the impurity solution containing the metal salt and both DDA and DDAB. Therefore, we suggest a doping mechanism where the impurities are initially introduced into the NC as Ag(I), followed by the reduction of the Ag(I) to Ag(0), by the DDA, on the surface of the NC once the "solubility limit" is reached.

The X-ray Absorption Near Edge Structure (XANES) spectra, as collected from the In and As K-edges for all three NC sizes (Fig. S12) reveals that the position of the absorption edge of both As and In edges do not change with $(Ag/NC)_i$ ratios, consistent with essentially unchanged charge states of both In and As.

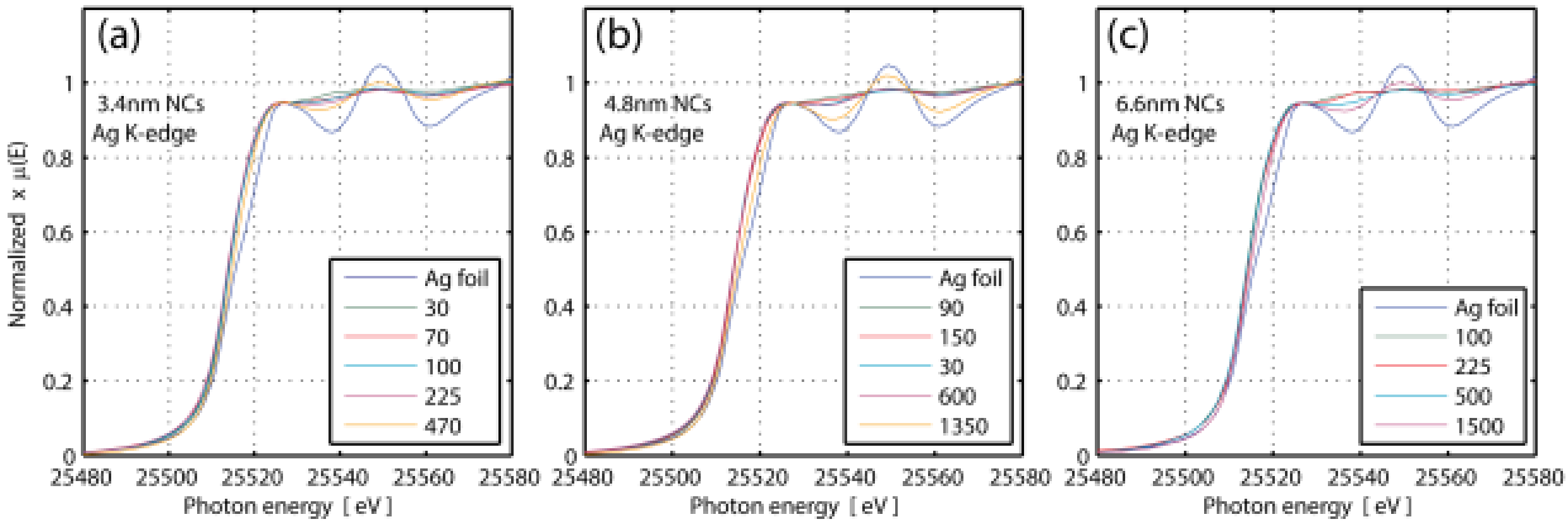

**Figure 5.** Ag K-edge XANES spectra for (a) 3.4nm, (b) 4.8nm, and (c) 6.6nm InAs NCs doped with Ag. The legend depicts the $(Ag/NC)_i$ ratios in the solution .

However, for low (Ag/NC)$_i$ ratios, the XANES spectra of the Ag K-edge (Fig. 5) as measured for 3.4nm, 4.8nm, and 6.6nm InAs NCs appears to be nearly featureless above the absorption edge. This is indicative of a significant disordering of the environment around the Ag atoms. Due to the ensemble-averaging nature of XAFS, it is impossible, using a measurement of a single sample or a single state of a given sample, to determine whether the disorder is large around each Ag atom, all of which are crystallographically equivalent, or if there is a heterogeneity in Ag placements in the InAs lattice, which causes the large disorder in the average configuration of neighbors around Ag. As the concentration of the Ag increases the fine structure above the absorption edge starts to be visible and for the highest Ag concentration of all three NC sizes, the structure of the spectrum resembles that of the metallic Ag reference, indicative of the formation of metallic state.

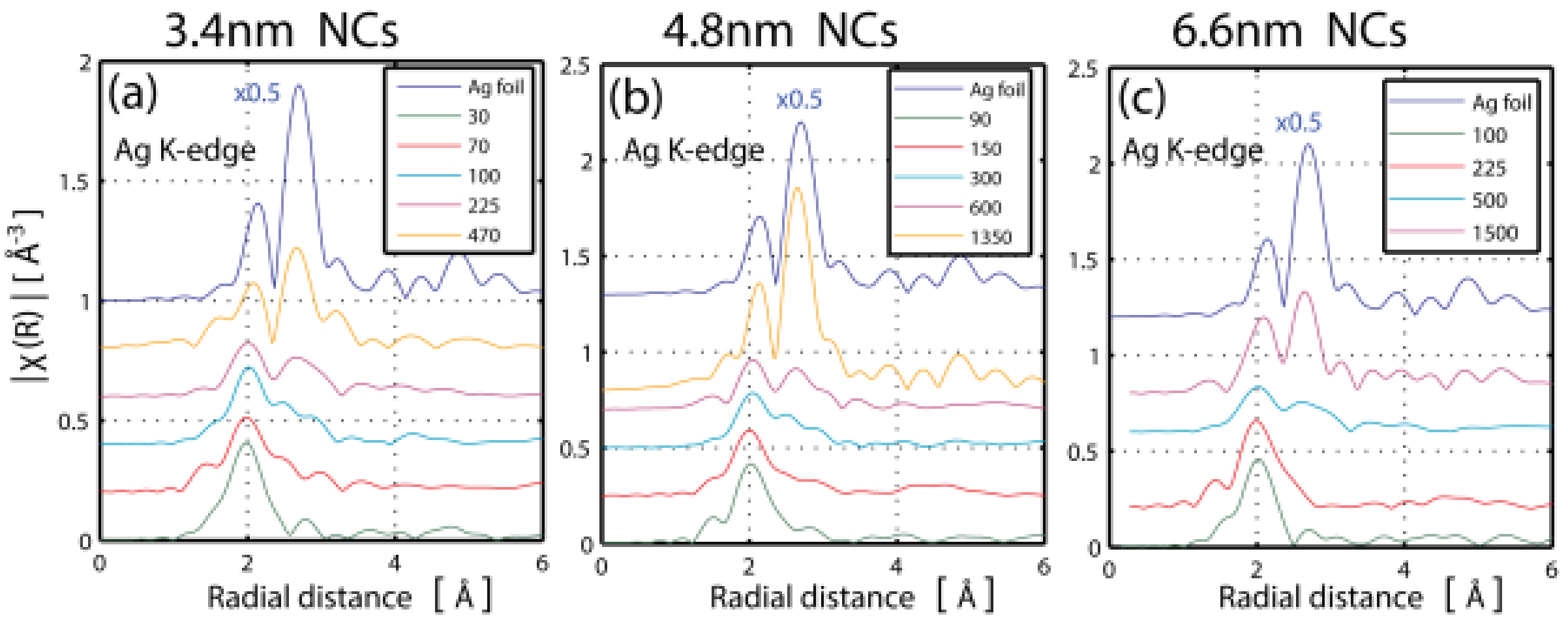

**Figure 6**. Fourier-transform magnitudes of the EXAFS spectra at different (Ag/NC)$_i$ ratios in solution for the Ag K-edge of the 3.4nm (a), 4.8nm (b), and 6.6nm (c) Ag doped InAs NCs. The spectra are offset vertically for clarity of presentation. The spectrum of the Ag reference foil is scaled down by a factor of 0.5. The legends depict the (Ag/NC)$_i$ ratios in solution.

Figure 6 shows the Fourier-transform magnitudes of the Extended X-ray Absorption Fine Structure (EXAFS) spectra for the 3.4nm (a), 4.8nm (b), and 6.6nm (c) Ag doped NCs measured from the Ag K-edge. The k-range used for the above Fourier transform was 2-12 Å$^{-1}$ for the In K-edge, 2-13 Å$^{-1}$ for the As K-edge and 2-11.4 Å$^{-1}$ for the Ag K-edge (for representative k-space spectra of the EXAFS data of the 4.8nm NCs see figure S13). Visual examination of the EXAFS data indicates that for both As and In edges (Fig. S14), the position of the main R-space peak is maintained and no new peaks evolve upon doping. This is consistent with substitutional doping, indicated also by the XANES spectra (Fig. S12) where Ag substitutes for In in the InAs lattice. This is in contrast to the behavior observed by us previously for Cu doped InAs NCs, which is an interstitial dopant and, accordingly, its incorporation in interstitial sites is directly evidenced by appearance of new peaks for the In and As in their EXAFS spectra and the formation of isosbestic points in both As and In XANES spectra.[38]

Unlike the In and As K-edges, the Ag K-edge data does change upon increasing the doping level. For low Ag levels, all the different sized NCs exhibit an apparent single R-space peak in the Ag EXAFS data, around R ~2 Å (Fig. 6a-c). This peak is distinctly different from the Ag-Ag spectra of pure Ag reference foil, and can be explained by Ag-As contribution only, indicating that at low $(Ag/NC)_i$ ratios no metallic Ag is formed in either samples. Increasing the $(Ag/NC)_i$ ratio beyond the "solubility limit", results in the appearance of a second peak around R ~3Å, in the region of Ag-Ag contribution of the bulk Ag. These changes, which will be confirmed by quantitative data analysis (*vide infra*), correspond to the formation of metallic Ag at higher Ag doping levels.

These observations for Ag are consistent with the following scenario: first, at low $(Ag/NC)_{i,}$ratios the preservation of EXAFS and XANES signals for the In and As edges is consistent with Ag doping, where Ag occupies substitutional (In) sites and not being at interstitial sites.[38] Indeed, *p*-type behavior was observed by us for Ag in InAs[31], consistent with substitutional doping. Additional support to this model is gained from the Ag K-edge EXAFS data behavior (Fig. 6a-c), dominated by the Ag-As contribution, as also evidenced by the quantitative data analysis. Thus, a structural model wherein Ag is a substitutional dopant was constructed and used to fit the data in this regime (Fig. S15).

Further details and insight to the Ag doping behavior in these low $(Ag/NC)_i$ ratios is obtained from the quantitative results (Table S2 and Fig. S16). The best-fit results consist of: (i) the coordination numbers (N) for the different neighbors surrounding the Ag atom, (ii) the inter-atomic distances (R) between the Ag and its nearest neighbors, (iii) the mean square deviations of the interatomic bond lengths, also known as EXAFS Debye-Waller factors ($\sigma^2$) for each bond, and (iv) the changes in the photoelectron energy origin ($\Delta E_0$).

As seen in the SI Table S2, the best fit results for the Ag-As coordination number (at low $(Ag/NC)_i$ ratios) yield an average coordination number closer to N=2, rather than to N=4 (Table S2). Attempting to constrain the Ag-As coordination number to be exactly N=4 (for the lower $(Ag/NC)_i$ ratios) resulted in $\Delta E_0$ values of -15eV for the Ag K-edge, which are non-physical. On the other hand, other contributions to Ag could be expected, such as Ag-N, due to the presence of amine-capping ligands. When added, the Ag-N contributions improved the fit reducing both values of reduced chi-square and R-factor (Table S3). Together, these two parameters (Ag-As and Ag-N coordination numbers), and the fact that Ag-N bonds were not detected in the solution, suggest that the Ag is situated predominantly on sites at the surface of the NCs. This model is different from that where Ag atoms are purely substitutional, *i.e.*, spread in the NC uniformly. In the latter model, the coordination number of Ag-As would be the same as In-As which are equal to 3.5, 3.7, and 3.7, for the particles of 3.4, 4.8 and 6.6 nm, respectively. Those numbers are in all cases larger than those obtained experimentally for Ag (Table S2).

However, sole surface binding of Ag atoms is inconsistent with the observed optical spectral red shift. We have previously modeled the contribution of surface strain to the evolution of the electronic properties of the InAs NCs upon doping, and found that purely surface bound impurities could not induce electronic shifts on the order of 100 meV.[31] This notion, together with the featureless XANES spectrum (Fig. 5) and the averaged coordination numbers (Table S2) for the low $(Ag/NC)_i$ ratios, suggest a more intricate explanation. The emerging behavior is one in which the Ag has in fact, a diverse heterogeneous surrounding, where some of the impurities are substitutional dopants and the rest are surface bound, occupying in particular As dangling bonds. This inhomogeneous distribution can result in such average behavior since the EXAFS method indeed provides the ensemble average data for the sample. This more elaborate view for Ag doping in InAs NCs is also in line with the intricate structure of Ag doping in CdSe studied by Norris and co-workers, which demonstrated the repulsion of Ag impurities to the surface of the NC where they assumed both interstitial and substitutional lattice sites.[32,39] However, unlike the system described by Norris *et. al.* attempts to allow for both Ag-As and Ag-In bonds (as would result from Ag being an interstitial impurity[38]) resulted in unphysically large $\Delta E_0$ for the Ag-In bond, indicating this type of binding is not favored by the fit.

Another interesting parameter obtained by the fit (Table S2) is the inter-atomic (R) distance between the probed atom and its surrounding neighbors. Figure 7 depicts the Ag-As (green) and In-As (black) inter atomic distance as a function of the $(Ag/NC)_a$ ratio for the 4.8nm doped InAs NCs (for the 3.4nm and 6.6nm sized NCs, see SI Fig. S17). As can be seen, the In-As bond length is unaffected by the doping for all three NC sizes, suggesting that the lattice structure is generally preserved during the doping process considering the large numbers of these original NC atoms (as can also be deduced from the small Debye-Waller factor, see SI Table S2). The Ag-As bond length, however, was found to expand in all three NC sizes when increasing the $(Ag/NC)_a$ ratio.

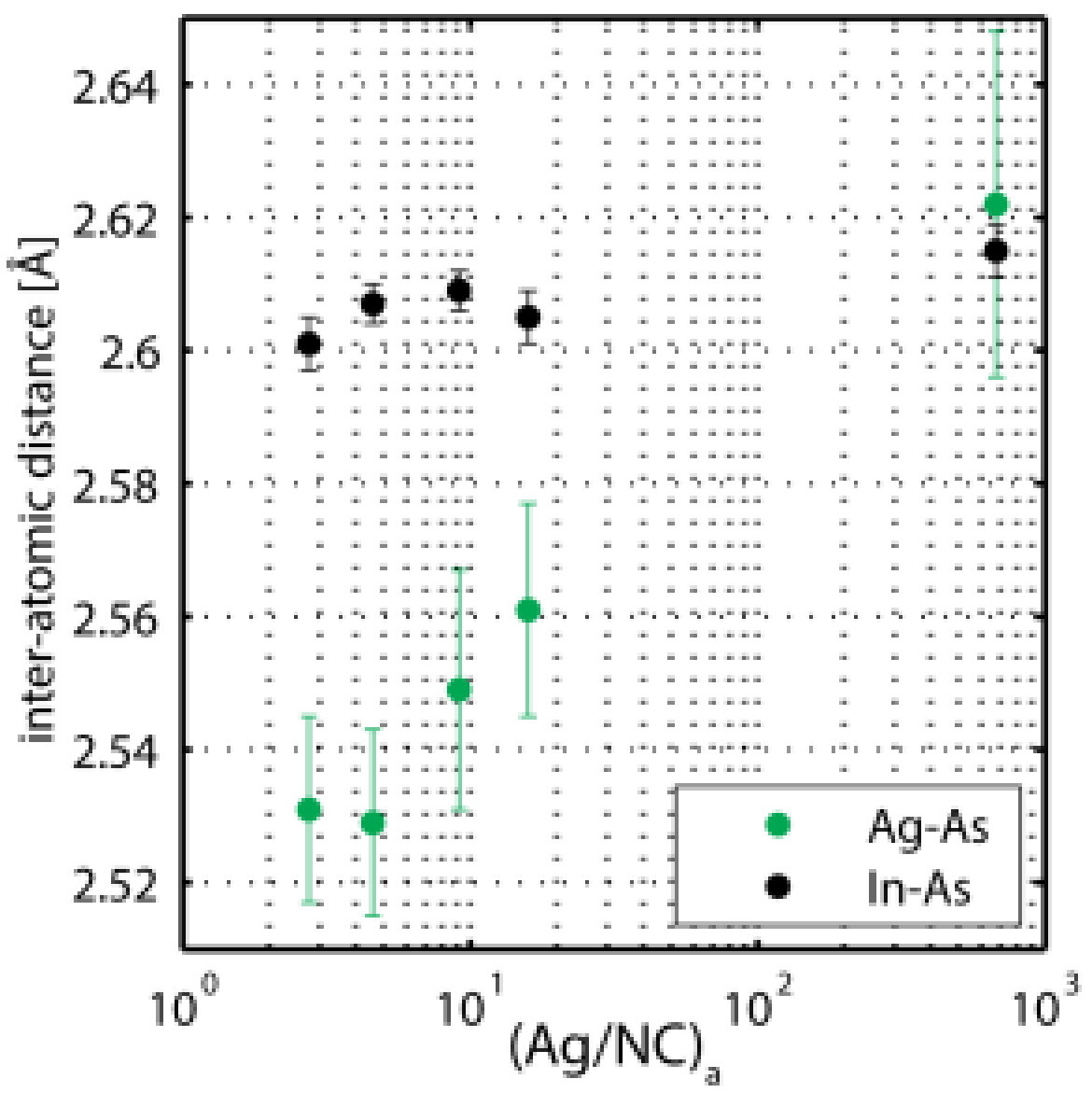

**Figure 7**. Ag-As (green) and In-As (black) inter-atoms distances extracted from the EXAFS fitting results, as a function of the $(Ag/NC)_a$ values for the 4.8nm doped InAs NCs.

As already discussed above, the solubility limit is of the order of few to tens of Ag atoms per NC. Interestingly, as Figure 7 demonstrates, for Ag doping levels under the solubility limit, the Ag-As bond length is shorter than that of In-As. The issue of impurity-host bond lengths in semiconductors has been discussed previously for isovalent impurities, suggesting that the host-impurity bond length is strongly affected by the size of the substituting atom, and the redistribution of charge density around the impurity-host bond.[43] Such changes in the impurity-host bond lengths were used in the case of doped ZnSe alloys to determine that the covalent radius of Co is slightly smaller than that of the Zn host atom within the crystal lattice. [44] More recently, a bimodal behavior was presented for Co doped CdSe and ZnSe NCs owing to the differences in the ionic radii of the constituents.[45] However, the differences in the ionic radii of the host and the impurity alone are insufficient to rationalize our experimental results (Fig. 7), which show a decrease in the Ag-As bond length at the impurity level (contradictory to the simple consideration where a larger impurity (Ag) should lead to a larger bond length when substituting for a smaller size host (In)). The explanation is in the change of effective ionic radius of As when Ag comes into its nearest neighbor position, compared to the undoped case. Such effect is possible due to the redistribution of the charge density around the Ag-As bond occurring upon doping in the InAs system, shifting the charge density from As-Ag towards As-In bonds. Due to the low level of Ag impurities, the effect on the In-As bonds is minimal and therefore no change is seen in the EXAFS results (Fig. 7). Furthermore, this interpretation also agrees with the electronic model of Ag doping, in which the Ag acts as an acceptor in the InAs lattice yielding a *p*-type NC.

We next discuss the second, growth regime, beyond the solubility limit of the Ag impurities. As interpreted in the visual examination of the EXAFS data (Fig. 6), an increase in the $(Ag/NC)_i$ ratio results in the formation of metallic characteristics, represented by the Ag-Ag interaction in the fit results (Table S2). The $(Ag/NC)_i$ ratios at which Ag-Ag bonds start to appear, increase with the size of the NCs, consistent with the previously discussed optical absorption and ICP-MS measurements. However, whereas all previous measurements (ICP-MS, optical absorption, and XANES analysis) indicate the strong accumulation of Ag occurs only beyond a $(Ag/NC)_i$ ratios of 200, 400, and 1000, for the 3.4nm, 4.8nm, and 6.6nm NCs, respectively, the EXAFS best-fit results indicate that some Ag-Ag bonds are present already at $(Ag/NC)_i$ ratios of 100, 300, and 500 Ag atoms for the respective sizes. This suggests that while Ag-Ag interactions start forming rather early, at low $(Ag/NC)_i$ ratios, the Ag clusters are not nucleated and formed, until the solubility limit is reached, after which these metallic phases act as a nucleation sites for rapid Ag accumulation.

The EXAFS best-fit results (Table S2) show that the highest doping levels for the 3.4nm, 4.8nm, and 6.6nm samples yield average Ag-Ag coordination numbers of 4.2±0.6, 7.4±0.6, and 5.0±0.8, respectively. However, considering the $(Ag/NC)_a$ values for these samples (175-1000 atoms), we have calculated the average coordination number and found it should be in the range of 7-10 (for detailed explanation see SI Fig. S18). There could be two different reasons for this discrepancy. The first is the formation of numerous ultra-small metal clusters (that lower the average metal-metal coordination number) on the surface of the NCs, and the second is the manner by which EXAFS obtains the coordination numbers in the case when two or more environments of Ag atoms are mixed together in the sample, *e.g.* substitutional Ag-As, surface bound Ag-As, and Ag-Ag.[46] We note in this regard that the inter-atomic distance of the Ag-Ag bonds, as indicated by Table S2, shows almost no change when increasing the Ag concentration in the solution phase. This differs from the expected trend for the growth of spherical-like metallic clusters, which are reported to have an initially reduced inter-atomic bond length compared to their bulk counterpart (2.87Å) and expand as the particle grows in size[47,48], further supporting the idea of the Ag having a heterogeneous surrounding and a non-homogenous metal growth on the surface of the NC.

This was further corroborated by powder XRD measurements that were performed on Ag doped InAs samples. XRD data of 4.8nm InAs NCs doped with different $(Ag/NC)_i$ ratios (Fig. 8) indicates that for low Ag concentration the main InAs reflections are preserved and no new peaks appear in the spectrum (see Fig. S19 for XRD spectra of 3.4nm and 6.6nm doped NCs). However, a new diffraction peak appears at 2θ=15 deg. (indicated by the arrow), identified as the (111) Ag crystal plane, upon reaching high doping concentrations. This indicates that while Ag-Ag interactions may be present already at low $(Ag/NC)_i$ ratios (as indicated by the EXAFS results), they are characterized by low crystallinity, and it is only past the "solubility limit" that metallic structures are formed on the surface of the NC, in agreement with all the previously interpreted data.

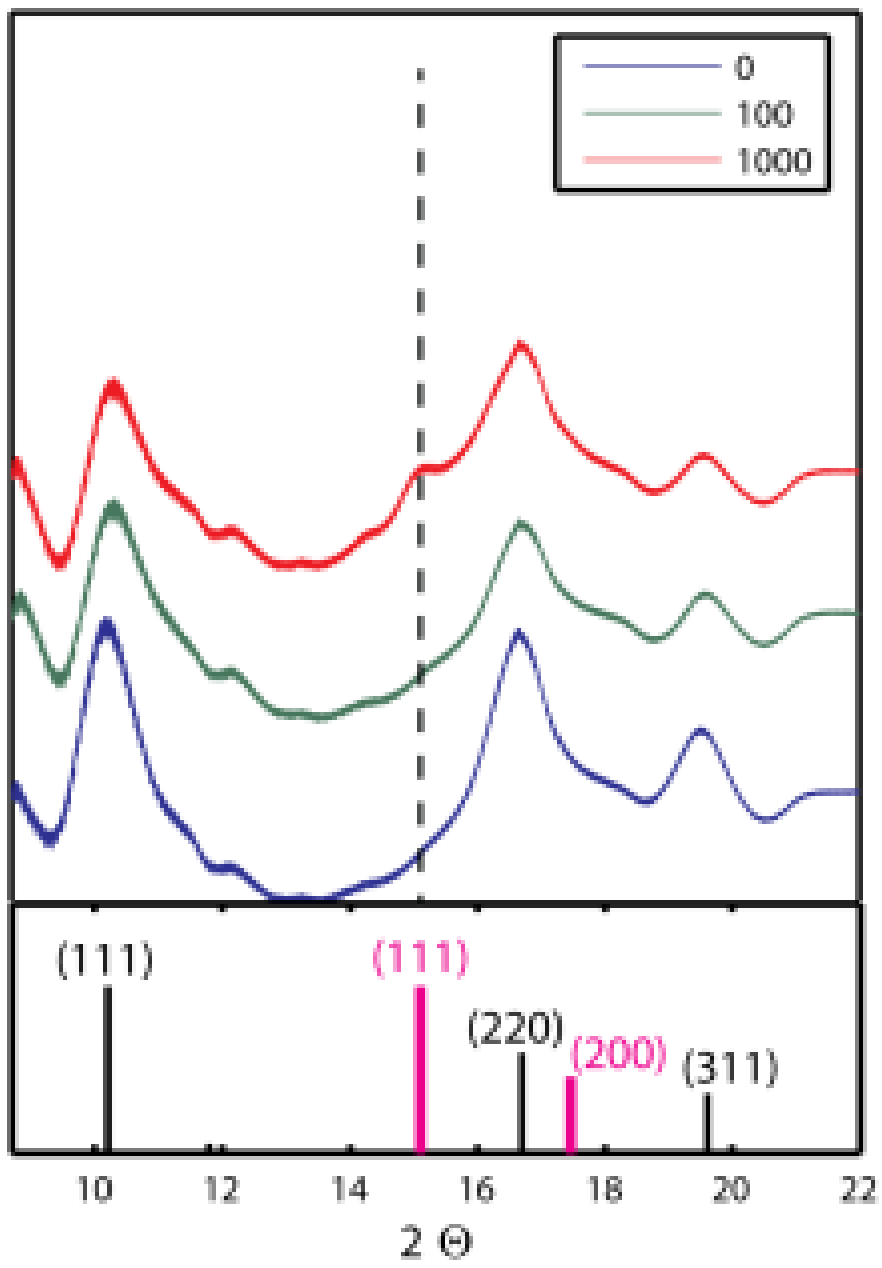

**Figure 8**. Synchrotron based powder XRD measurements (λ=0.62Å) for 4.8nm InAs NCs doped to various $(Ag/NC)_i$ ratios. Bottom panes represent the standard diffraction peaks for InAs zincblende lattice (black bars) and Metallic Ag FCC (magenta bars). The dashed line represents the position of the (111) Ag crystal plane.

**Conclusions.** A study of diffusion-based Ag doping in pre-synthesized InAs NCs of different sizes was carried out. Optical absorption and ICP-MS measurements indicated two different regions that depend on the $(Ag/NC)_i$ ratio in the solution: (i) the "doping region" at low $(Ag/NC)_i$ ratios and the "growth region" at high ratios. Combination of advanced XAS and XRD measurements resulted in a structural model of Ag(I) initially substituting for In atoms near the surface of the NC followed by the formation of Ag metallic structures on the NC surface. As the $(Ag/NC)_i$ ratio increases, larger metallic clusters begin to form, acting as nucleation sites and preventing additional doping of the NC, which has already reached the Ag solubility limit. The results presented here provide insight towards the "doping limit" of colloidal semiconductor

NCs, which has not been observed in previous diffusion based doping reactions such as Cu doping in InAs[38], Ag doping in CdSe[32], or Mn doping in CdSe.[33] This kind of understanding of various behaviors of semiconductor nanocrystals doping scenarios is a necessary prerequisite for rational design of doped nanocrystals with desired properties. Availability of such doped nanocrystals may open a path for further development of advanced optoelectronic devices based on heavily doped semiconductor NCs.

## Experimental methods

**Colloidal InAs NCs were synthesized** following a well-established wet-chemical synthesis.[40] Precursor solutions containing $(TMS)_3As$ (Tris(trimethylsilyl)arsine) and $InCl_3$ (Indium(III) chloride) were prepared in a nitrogen glove-box (MBraun) and kept under inert conditions throughout the reaction. A solution of distilled TOP (Trioctylphosphine) was evacuated for ~30 min and then heated to 300°C. The nucleation solution was rapidly injected and the solution temperature lowered to 260°C. The growth solution was then gradually introduced to the solution, allowing particle growth until the desired size is reached. Narrow size distributions were further achieved through size selective precipitation performed in a glove-box by adding methanol to the NC dispersion and filtering the solution through a 0.2 um polyamide membrane filter (Whatman).

**Doping of pre-synthesized InAs NCs** was achieved *via* a room-temperature solution-phase reaction previously reported by us.[31] Briefly, an impurity Ag stock solution was prepared by dissolving 10 mg (0.058mmol) of the metal salt ($AgNO_3$), 80mg (0.17mmol) Didodecyldimethylammonium bromide (DDAB), and 120mg (0.64mmol) Dodecylamine (DDA) in 10ml of anhydrous-toluene (Sigma). Calculated amounts (V/V) of the impurity solution were added to a suspension of InAs NCs, according to the desired $(Ag/NC)_i$ ratio, while stirring. The concentration of the NCs was estimated based on the literature values of the InAs NC absorption cross-section.[49] The doping reaction was performed under inert condition in a glove-box and the reaction was terminated after 15min by adding methanol and isolating the doped NCs through precipitation.

**ICP-MS measurements** (Inductively coupled plasma – mass spectroscopy) were carried out in an Agilent 7500cx in order to correlate between the solution ratio and the actual number of impurities per NC. First, the NC samples were dried under vacuum, after which they were dissolved in a 70% $HNO_3$ solution (Sigma) and diluted to a final concentration of 3% $HNO_3$ with triply distilled water (TDW). The impurity concentration was calculated *versus* a calibration curve based on reference samples prepared using a 1000ppm standard solution (High Purity Standards). Prior to the analysis, the ICP-MS was calibrated with a series of multi-element standard solutions (1ng/l—100 mg/l Merck ME VI) and standards of major metals (300mg/l—3 mg/l). Internal standard (50 ng/ml Sc, 5 ng/ml Re and Rh) was added to every standard and sample for drift correction.

**XAFS measurements** were performed at the Brookhaven National Laboratory (BNL) National Synchrotron Light Source (NSLS) beamline X18B. The specimens were prepared under inert conditions in a glove-box (MBraun) where they were spread on adhesive tape and mounted in a sample chamber. The chamber, filled with nitrogen, was then transferred to the beamline where it was flushed with nitrogen during the measurements to maintain inert conditions through the experiments. Experiments at the As and In K-edges were performed in transmission mode, and for Ag K-edge – both in transmission and fluorescence. Three to ten consecutive measurements were taken at each edge, to improve the signal to noise ratio in the data.

**XAFS data analysis and processing** was done using standard techniques. Briefly, the data were first aligned in energy, using reference spectra collected in standard materials together simultaneously with the NC data, and then averaged. Next, the smooth, isolated atom background function was subtracted from the absorption coefficient using Athena program[50] from the IFEFFIT package.[51]

In all fits, theoretical photoelectron scattering amplitudes and phases were calculated using program FEFF6.[52] With the purpose of improving the reliability of the fit, for each sample size and each absorption edge, the fitting process was performed simultaneously for the different doping levels. For Ag K-edge EXAFS fits, Ag-In, Ag-N and/or Ag-Ag paths were included in the fitting model. The reduction factor ($S_0^2$=0.95) was found from the fit to the bulk Ag and fixed in the fits to the doped NC data. Adjustable parameters in the theoretical EXAFS signal included the coordination numbers of nearest neighbors of a certain type around the absorbing Ag atom (N), the bond distance between the absorber and the nearest neighbors (R), and its mean-square-displacement ($\sigma^2$). While these three variables (N, R and $\sigma^2$) changes with the doping levels, the corrections to the photoelectron energy origins ($\Delta E_0$) were varied independently for each neighboring pair but constrained to be the same for all doping concentrations of a certain sized sample. For In and As K-edge EXAFS fits, only one path was used for fitting. It is In-As for In edge data and As-In/Ag for As edge (the photoelectron scattering features of In and Ag are similar and cannot be discriminated in XAFS fitting), and their coordination number was constrained to be 4. Similar to the strategy used for fitting Ag edge data, R and $\sigma^2$ were global variables while $\Delta E_0$ was constrained to be the same for all doping levels of each sized sample. The reduction factors ($S_0^2$=0.89 for both In and As) for In and As edge data were obtained by fitting InAs QDs. The fitting range in k space is 2-11.4 Å$^{-1}$ for Ag, 2-12 $Å^{-1}$ for In and 2-13 $Å^{-1}$ for As. The fitting range in R space is 1.4-3.0 Å for Ag, 1.8-2.9 Å for In and 1.8-2.8 Å for As. For Ag edge spectra, the fitting k weight is 2 while 3 for fitting both In and As edge spectra.

**XRD measurements** were performed at the Brookhaven National Laboratory (BNL) National Synchrotron Light Source (NSLS) beamline X18A. The specimens were prepared under inert conditions in a glove-box (MBraun) where they were spread on adhesive tape. XRD patterns were acquired with a Perkin-Elmer (PE) amorphous silicon detector with 2048 × 2048 pixels and a 200 × 200 $\mu m^2$ pixel size. The detector is mounted on the 2θ arm, the typical 2θ range is 5 to 22°. The energy of the beam was set to 20 keV. Before spectra collection, the detector was calibrated using a $LaB_6$ standard. For each spectrum, multiple scans were recorded for both the diffraction pattern and the dark current, which was subtracted.

**Supporting Information Available:**

UV-VIS absorption spectra and ICP-MS measurements for additional NC sizes doped with various impurity concentrations. XAS As K-edge and In K-edge spectra for additional NC sizes as well as impurity solution measurements. Structural model used for fitting EXAFS data, best-fit results for all NC sizes and $(Ag/NC)_i$ ratios, and theoretical calculations of ideal metallic clusters coordination-numbers. This material is available free of charge *via* the Internet at http://pubs.acs.org

**Acknowledgements:**

The research leading to these results has received funding through the NSF-BSF International Collaboration in Chemistry program. YA, AIF and UB acknowledge the support of this work by the NSF Grant No. CHE-1413937 and the BSF Grant No. 2013/610. The work at X18B and X18A beamlines was supported, in part, by Synchrotron Catalysis Consortium (U. S. DOE Grant No. DE-FG02-05ER15688). We thank Steve Ehrlich for his help with XRD measurements at X18A and Nebojsa Marinkovic for his help with the XAFS measurements at the X18B beamlines. UB thanks the Alfred and Erica Larisch Memorial Chair.

# Supporting Online Material:

From Impurity Doping to Metallic Growth in Diffusion Doping: Properties and Structure of Ag Doped InAs Nanocrystals

*Yorai Amit*[1,2], *Yuanyuan Li*[3], *Anatoly I. Frenkel*[*,3] *and Uri Banin*[*,1,2]

[1]The Institute of Chemistry, Hebrew University, Jerusalem 91904, Israel.

[2] The Center for Nanoscience and Nanotechnology, Hebrew University, Jerusalem 91904, Israel.

[3]Department of Physics, Yeshiva University, New York, New York 10016, United States

* Address correspondence to: anatoly.frenkel@yu.edu, uri.banin@mail.huji.ac.il

In order to investigate the role of each constituent in the doping process, and reveal the mechanism of impurity doping, three precursor solutions were prepared: (i) 10mg $AgNO_3$ dissolved in 10mL of DDAB (0.017M), (ii) 10mg $AgNO_3$ dissolved in 10mL of DDA (0.064M), and (iii) 10mg $AgNO_3$ dissolved in 10mL of a DDA+DDAB solution (0.017M and 0.064M, respectively). After several hours, at room-temperature, under inert-conditions, the $AgNO_3$ in DDA solution changed its appearance from colorless to pale brown, while both $AgNO_3$ in DDAB and $AgNO_3$ in DDA+DDAB solutions remained colorless (Fig. S1b).

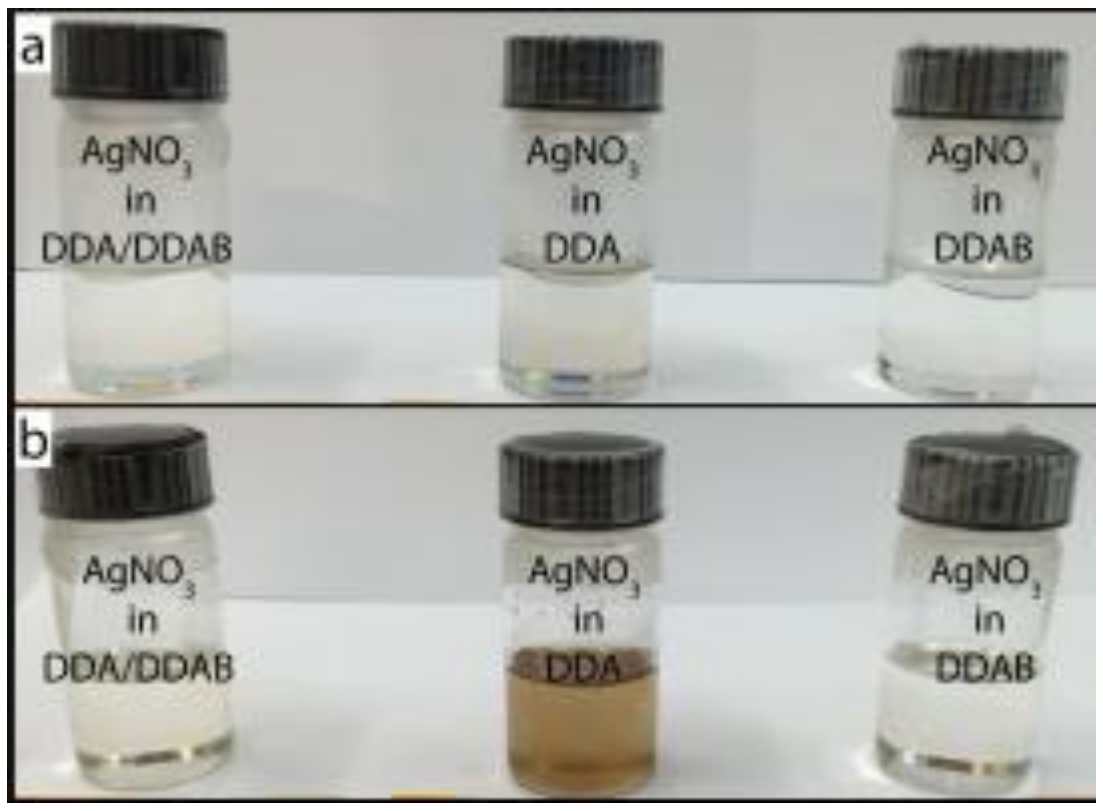

**Figure S1.** Pictures of (a) $AgNO_3$ dissolved in different ligand compositions, and (b) after 16 hours kept in a nitrogen glove-box at room temperature.

To prove the formation of metallic Ag, the product of the $AgNO_3$ in DDA solution was imaged by TEM and measured using powder XRD (Fig. S2). TEM imaging revealed the formation of aggregates and Ag nanoparticles (Fig. S2a). XRD measurements (Fig. S2b) revealed metallic Ag reflections together with the signature of the excess DDA. A $Ag_2O_3$ phase is also observed in the XRD due to exposure of the sample to air during the measurements.

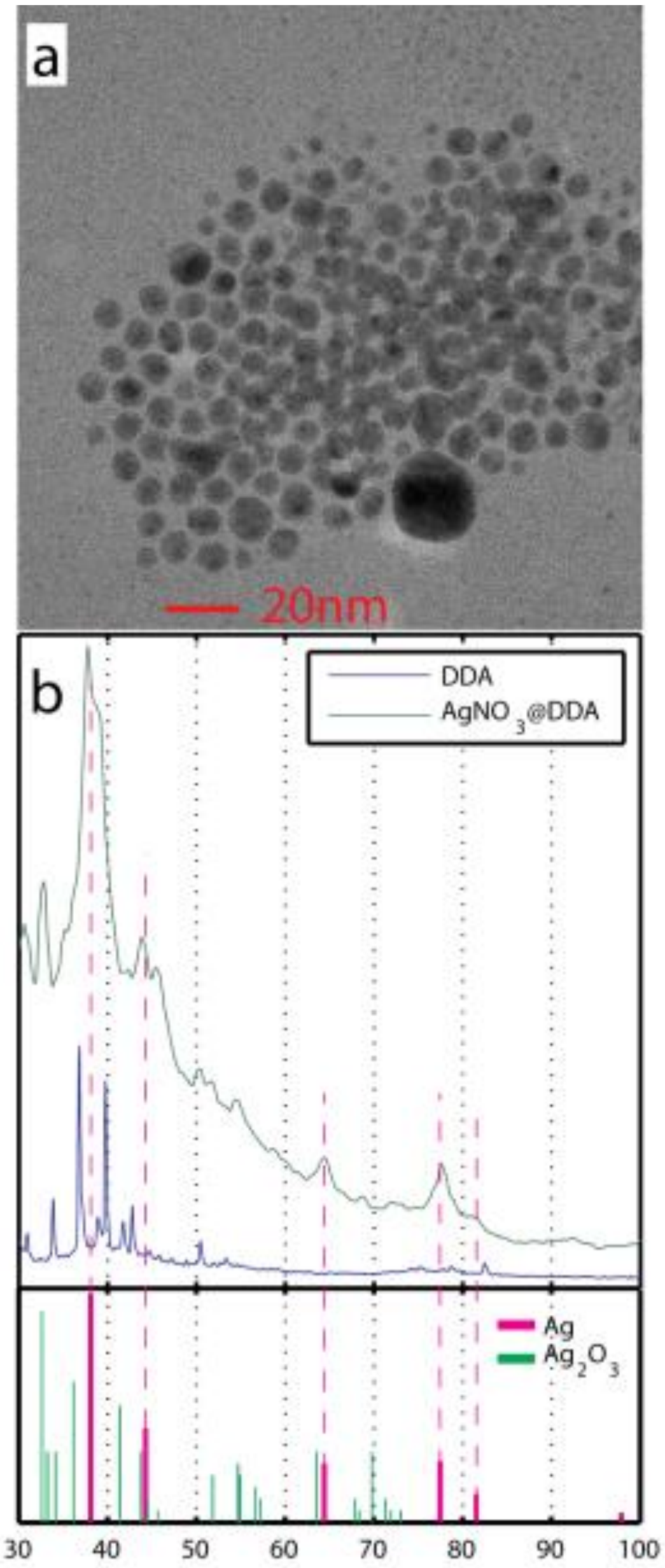

**Figure S2.** (a) TEM image of the product of the $AgNO_3$@DDA solution after 16 hours and (b) XRD diffractogram of the resulting product, indicating the formation of metallic Ag nanoparticles.

Figure S3 presents size histograms calculated using the TEM images presented in the main manuscript. >200 particles were measured for each sample to determine the size distribution. While the control sample has a mean diameter of 4.8nm and a standard deviation (SD) of 0.08nm both Ag doped samples maintain the mean diameter (4.8nm) and exhibit similar SD values (0.11nm and 0.08nm for the 400 $(Ag/NC)_i$ and 600 $(Ag/NC)_i$ in solution, respectively). To account for the observed optical shifts observed upon doping to these levels (>500 meV) the mean diameter of the NCs should have increased by over ~0.3nm. Therefore, we rule out any structural changes and conclude that the observed spectral shifts are a result of doping.

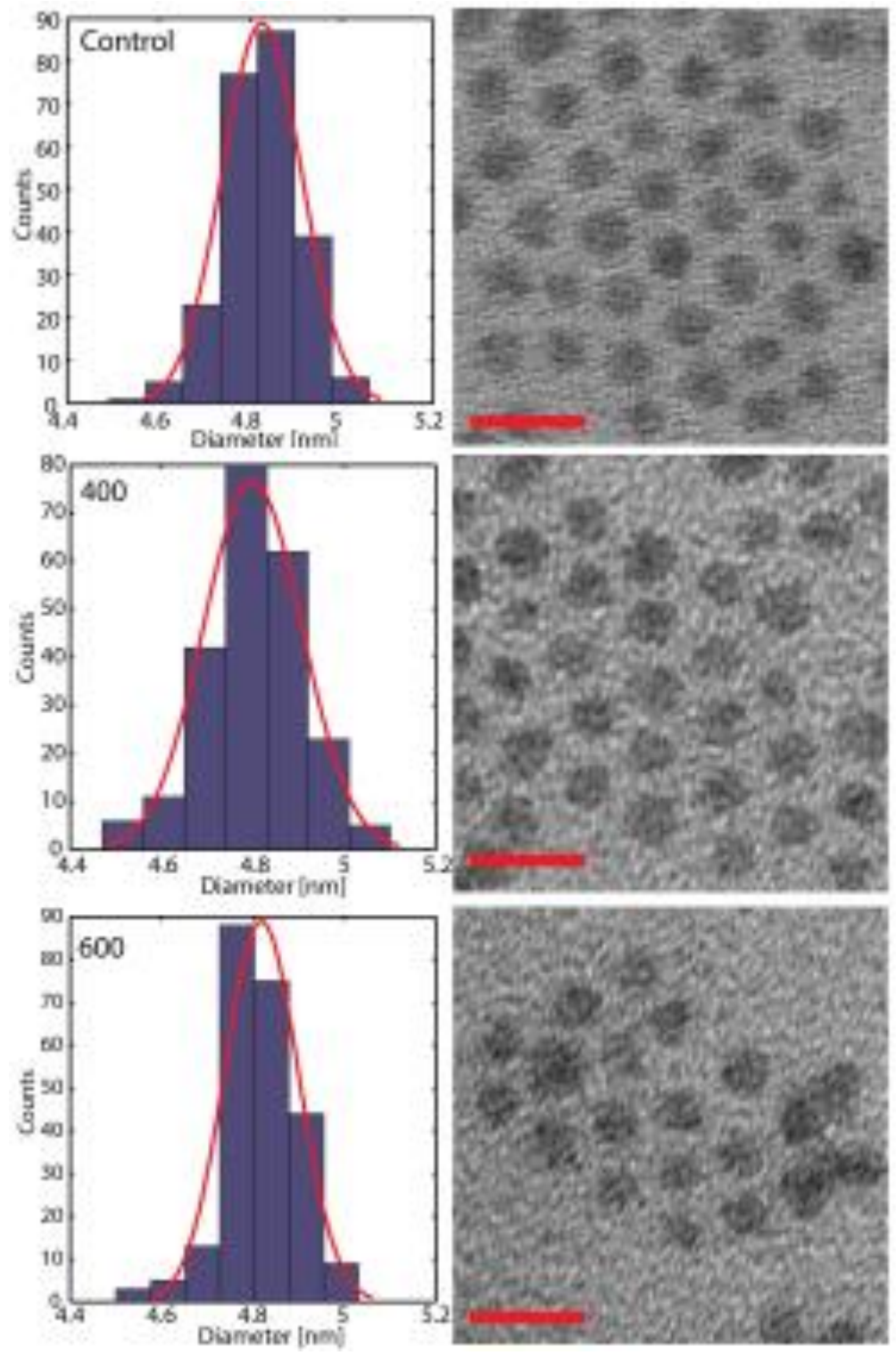

**Figure S3**. Size histograms (left) and corresponding TEM images (right) of three InAs NC samples. (top) Control sample. Mean diameter 4.8nm, standard deviation (SD) 0.08nm. (middle) 400 $(Ag/NC)_i$ in solution. Mean diameter 4.8nm, SD 0.11nm. (bottom) 600 $(Ag/NC)_i$ in solution. Mean diameter 4.8nm, SD 0.08nm. TEM scale bar is 10nm.

Figure S4 presents the optical absorption spectra for the (a) 3.4nm and (b) 6.6nm InAs NCs showing the evolution of the optoelectronic properties for various $(Ag/NC)_i$ ratios. As can be seen, the position of the first exciton peak gradually red-shifts[1] when increasing the $(Ag/NC)_i$ ratio. This is in accordance with previous STM studies, optical spectroscopy measurements, and theoretical analysis, which showed that the doping of InAs NCs with Ag impurities leads to *p*-type behavior and the red shift is related to band tailing effects associated to the disorder introduced by the impurity insertion.[2]

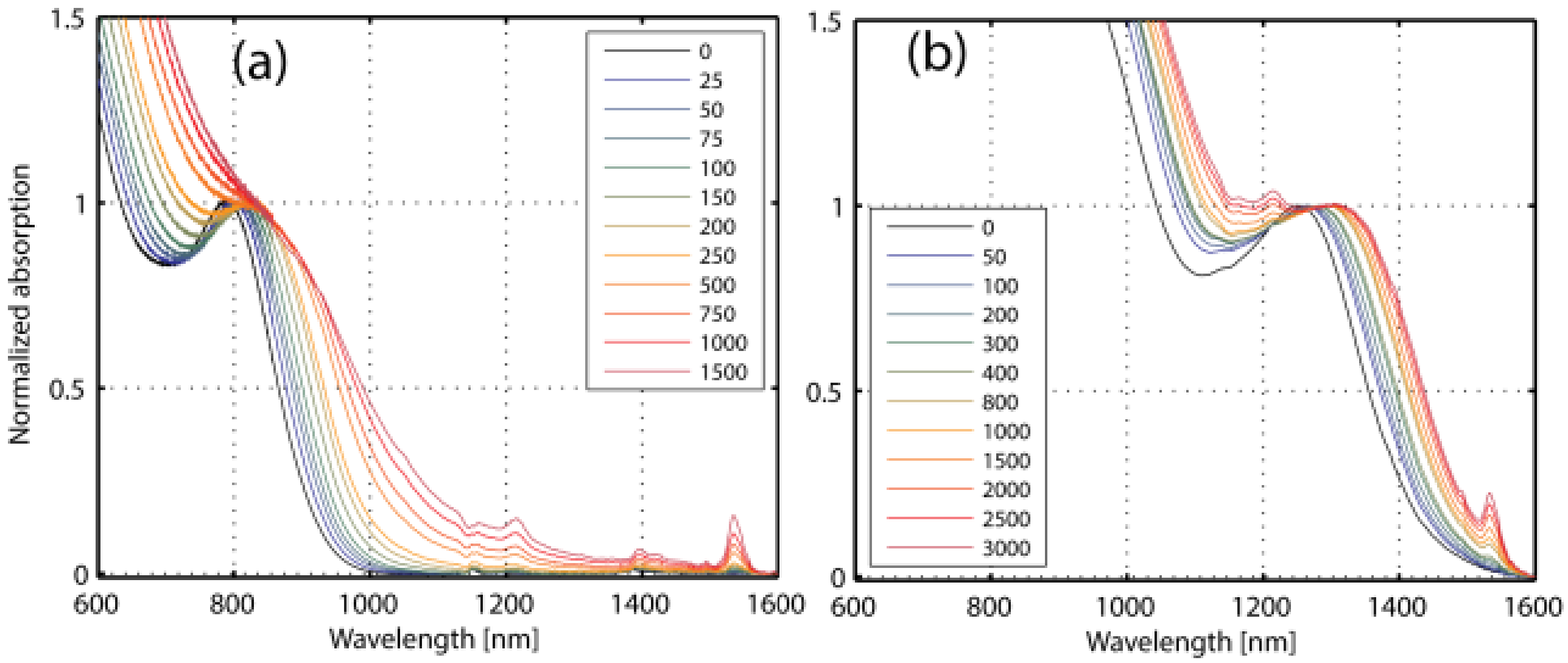

**Figure S4**. Optical absorption spectra of (a) 3.4nm InAs NCs and (b) 6.6nm InAs NCs for various doping levels of $(Ag/NC)_i$ ratios, which denote the ratios of Ag to NC in solution phase.

Measuring the optical absorption of the ligand solution reveals two dominant absorption peaks at ~1200nm and ~1600nm, rationalizing the emerging features in the optical absorption of the doped NC samples, at high doping ratios.

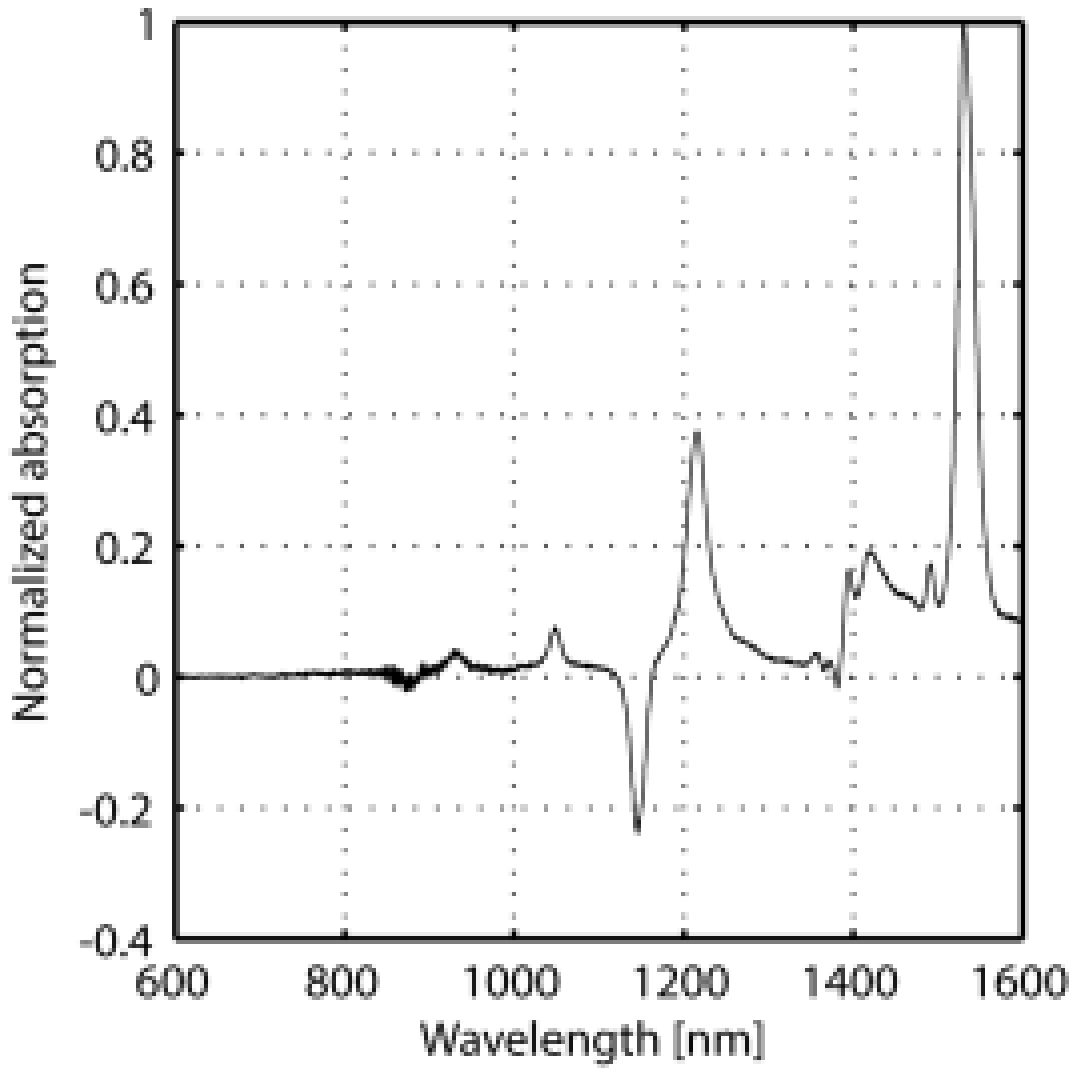

**Figure S5**. Optical absorption spectra of the control solution containing DDA and DDAB

Figure S6 shows the optical absorption spectra of 3.5 nm InAs NCs before and after doping with Ag for various NC and impurity concentrations. Throughout this experiment the total number of NCs in the solution was kept constant and only the concentration of the solution was varied (Fig, S6a). Two series of such NC solution were prepared and each set was doped with a constant amount of Ag impurities. Figure S3b shows the absorption spectra of the NC solutions after doping with 7.4umol of Ag (corresponding to 90 $(Ag/NC)_i$ in solution). Figure S6c shows the absorption spectra of the NC solutions after doping with 14.8umol of Ag (corresponding to 170 $(Ag/NC)_i$ in solution). Figure S6d shows that the key parameter controlling the doping reaction is in fact the Ag to NC ratio, as each of the two NC series exhibits a constant optical shift independent of the Ag concentration.

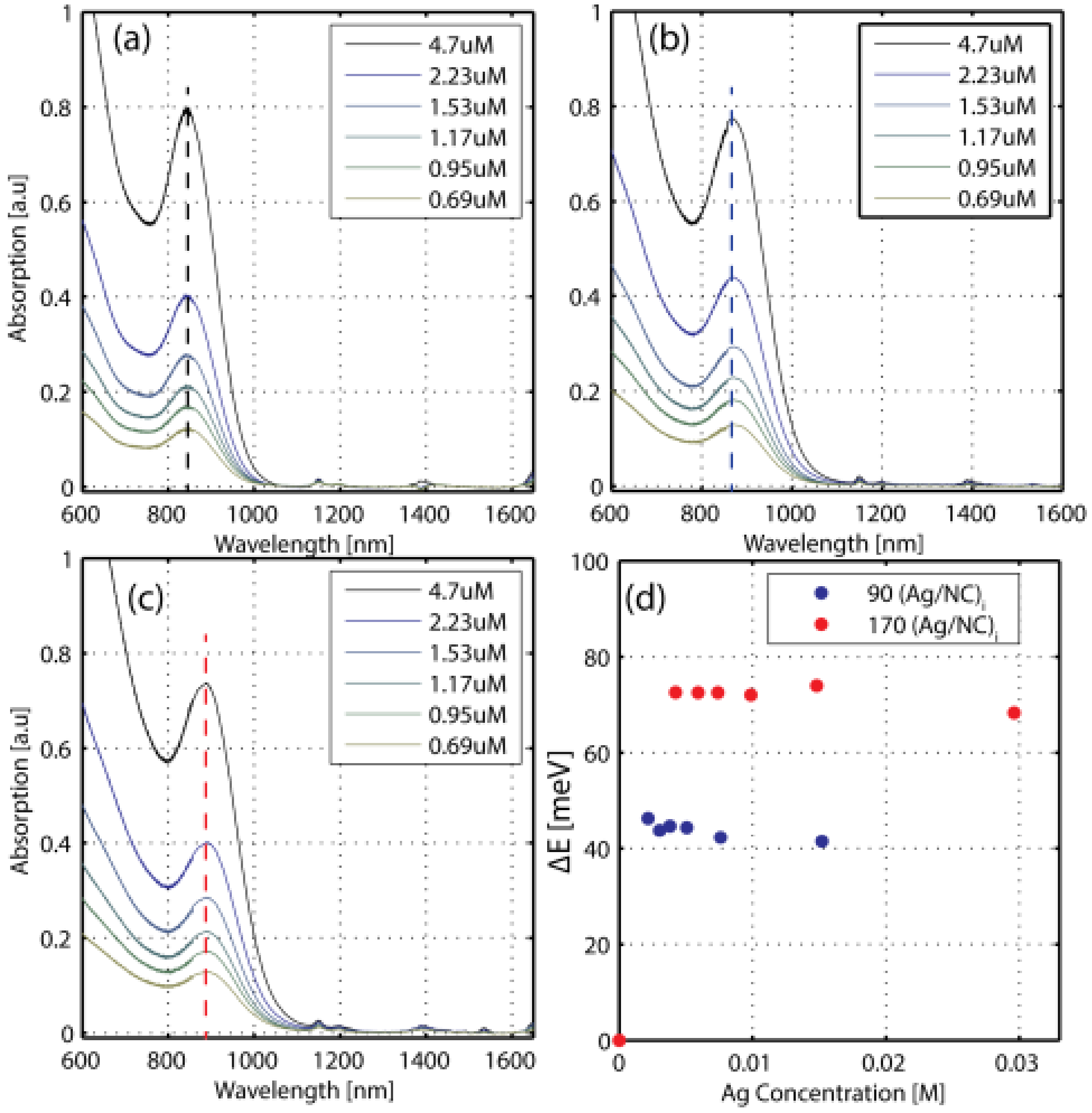

**Figure S6:** Optical absorption spectra for 3.5nm InAs NCs (a) before, (b) doped with 90 (Ag/NC)$_i$ in solution, and (c) 170 (Ag/NC)$_i$ in solution. (d) shows the resulting optical shift compared to the un-doped sample as a function of the Ag concentration in the solution.

Figure S7 shows the ICP-MS measurements of the 3.4nm, 4.8nm and 6.6nm InAs NCs doped with various $(Ag/NC)_i$ ratios in solution (Fig. S4 a-c, respectively). As can be seen, the data can be divided into two distinct regions: The "doping region", which is shown in the inset of each frame, is characterized by low yields of Ag atoms per NC. The "growth region", which occurs after crossing a critical $(Ag/NC)_i$ threshold is characterized by a strong accumulation of Ag atoms. The intersection point between these two regions is therefore considered as the "solubility limit". By linear fitting of each portion of the ICP-MS spectra, and calculating the intersection point of the two, we were able to estimate both the "solubility limit" in terms of $(Ag/NC)_a$ ratios, which, due to the low levels detected, is subject to large uncertainty values, and the "solubility limit" in terms of the $(Ag/NC)_i$ ratios in the solution.

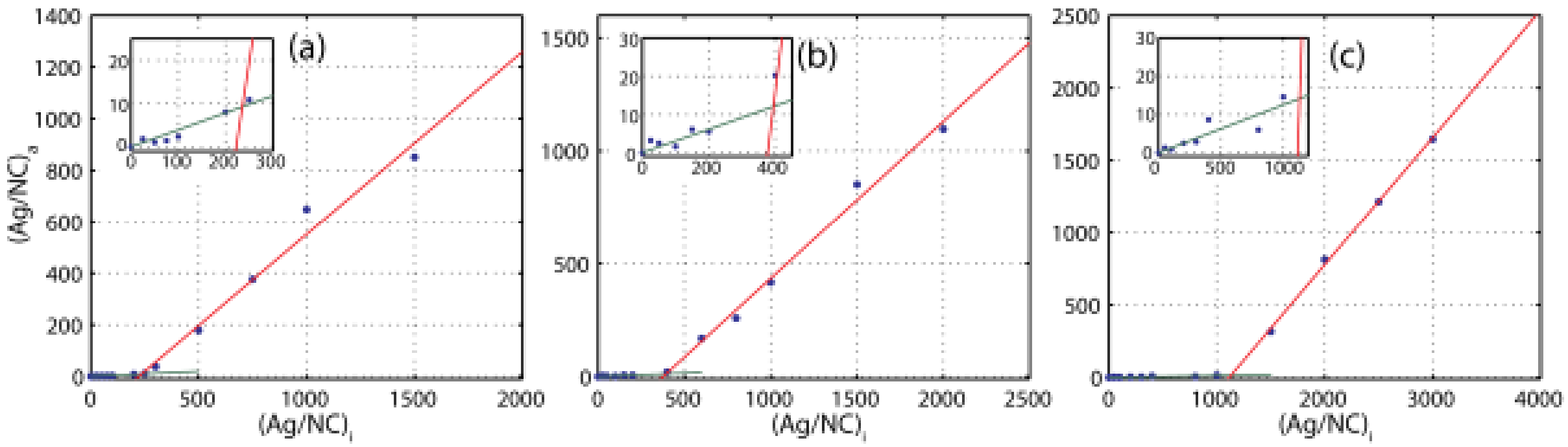

**Figure S7**. Linear fits for the different regions in the $(Ag/NC)_a$ ratio (obtained by ICP-MS) as a function the $(Ag/NC)_i$ ratio in solution for the (a) 3.4nm, (b) 4.8nm, and (c) 6.6nm InAs NCs doped with Ag.

Figure S8 shows the fitting attempts of the $(Ag/NC)_i$ "solubility limit" values, with the radius of the NC (a), the surface area of the NC (b), and once again the volume of the NC (c, similar to Fig. 3 in the main paper).

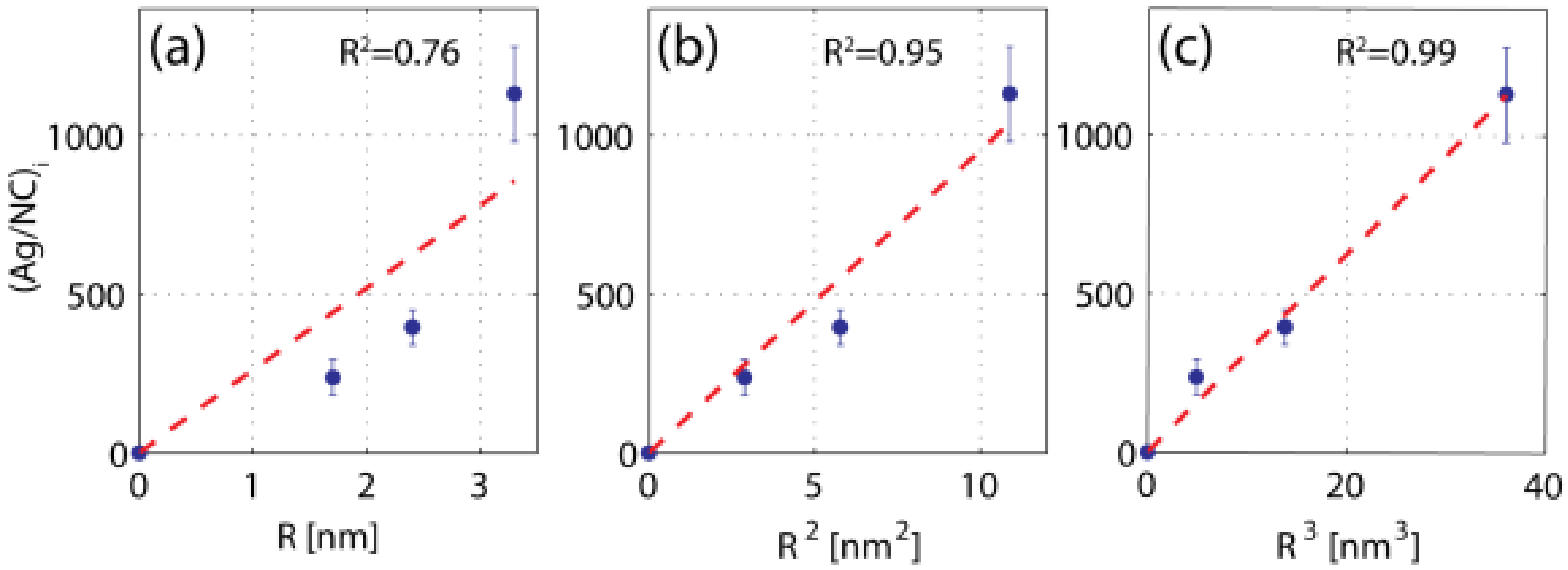

**Figure S8**. Linear fits for the "solution solubility limit" Vs the NC radius (a) NC surface area (b) and the NC volume (c). The goodness of the fit is denoted by the corresponding $R^2$ values.

Figure S9 depicts different scaling attempts between $(Ag/NC)_a$ ratios at the crossover point, and the size of the NCs. As can be seen, given the large uncertainties, no clear correlation can be identified.

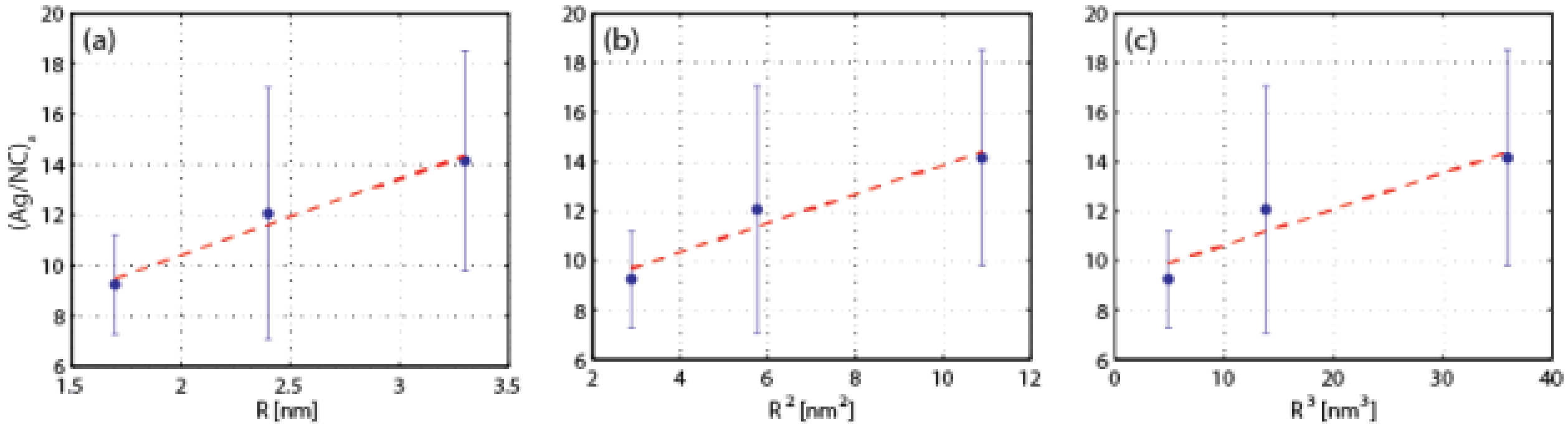

**Figure S9**. Extracted "solid solubility limits" as a function of (a) NC radius, (b) Surface area, and (c) NC volume.

In order to determine the mechanism by which the doping reaction takes place we have measured the XAS of the Ag doping solution (figure S10). From the best-fit results of this sample (Table S1) we conclude that the $AgNO_3$ has completely dissociated in the solution and all Ag atoms are coordinated with Br atom from the DDAB ligands present in the solution.

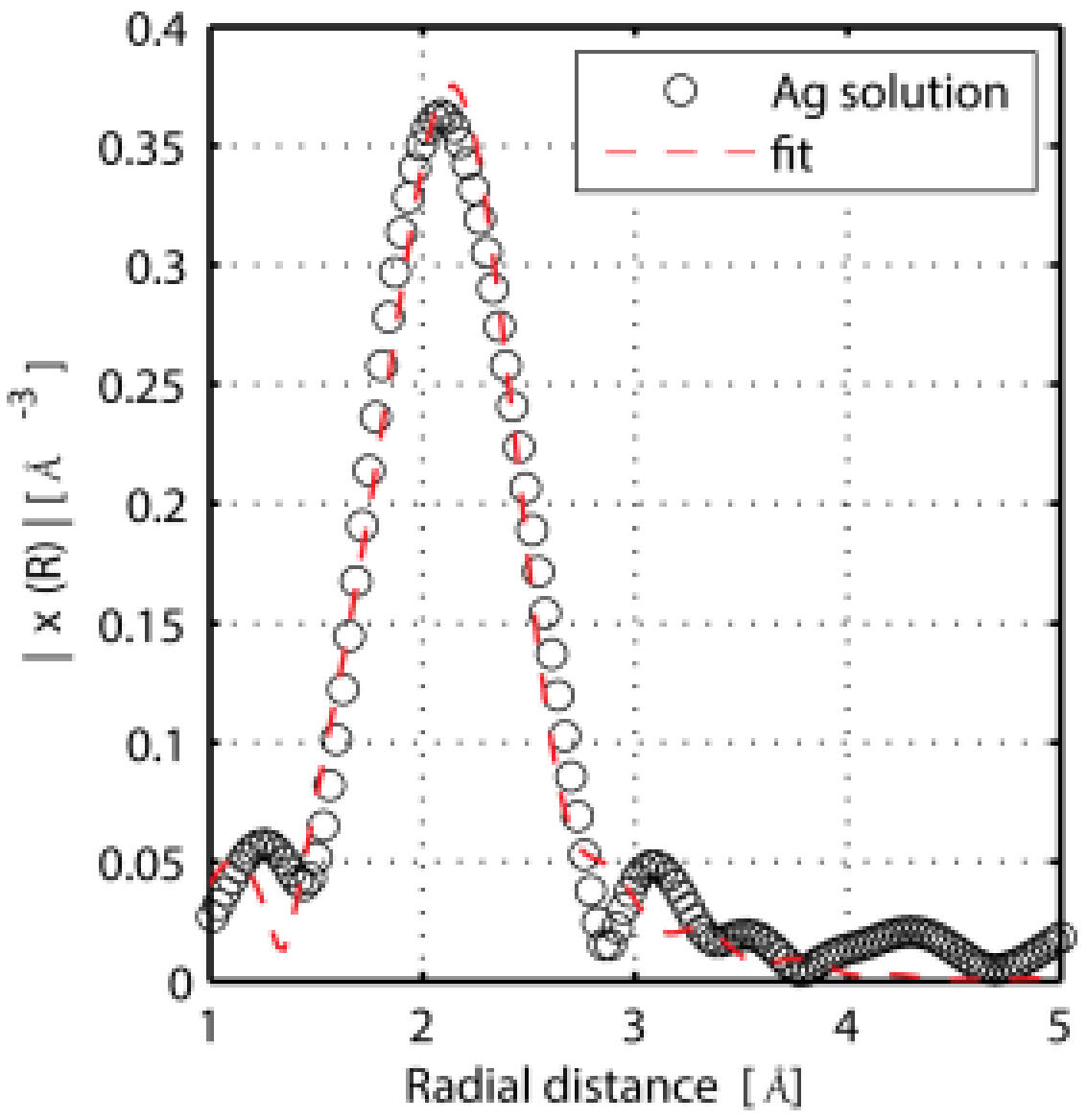

**Figure S10**. Fourier-transform magnitudes of the Ag doping solution Ag K-edge EXAFS spectra (o) and best-fit results (dashed line)

**Table S1**. Best-Fit Results for Structural Parameters of the Ag-Br Nearest Neighbor Pairs Obtained by Ag K-Edge Analysis of the Ag Doping solution

| Parameter | Value |
|---|---|
| $\Delta E_0$ [eV] | -5.75±2.73 |
| $N_{(Ag\text{-}Br)}$ | 2.2±0.6 |
| R [Å] | 2.54±0.15 |
| $\sigma^2$ [Å$^2$] | 0.007±0.002 |

To investigate the role of each constituent in the doping solution three different doping solutions were used and the optical absorption spectra compared for similar $(Ag/NC)_i$ ratios (Fig. S11). Three 3.4nm InAs NC solutions ($2.5*10^{-9}$M) were doped to a ratio of 500 $(Ag/NC)_i$ using either $AgNO_3$@DDAB, $AgNO_3$@DDA, or $AgNO_3$@DDA/DDAB. As seen in Figure S8, both $AgNO_3$@DDAB and $AgNO_3$@DDA+DDAB resulted in similar overall optical shift whereas the $AgNO_3$@DDA resulted in the complete loss of the absorption feature (as seen for higher doping ratios (>1000 $(Ag/NC)_i$), see Fig. S4a).

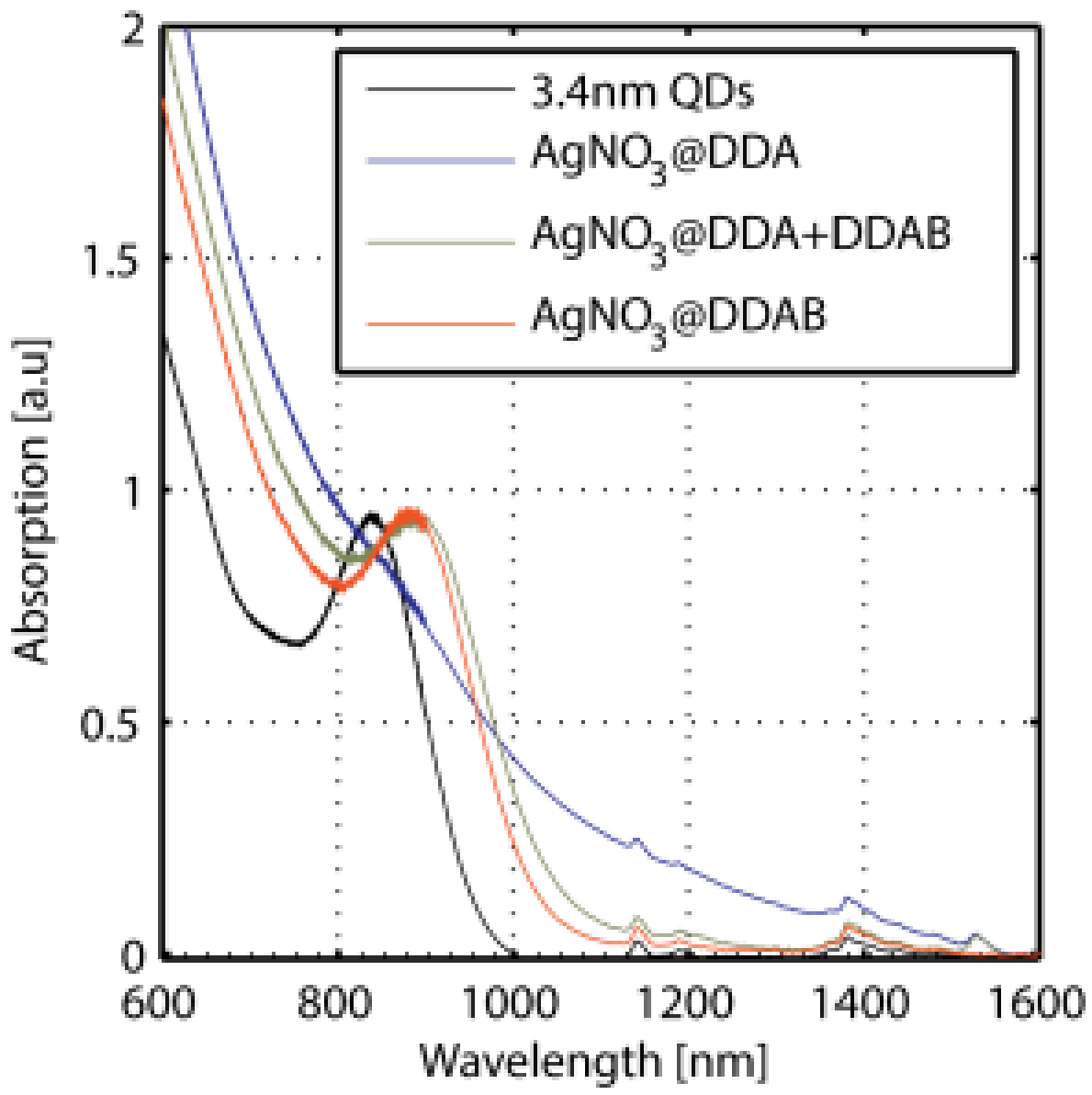

**Figure S11.** Comparison of the doping effects on the optical absorption spectrum when using different precursor solution. Red and Green curves show the optical absorption of 3.4 nm InAs NCs doped with 500 $(Ag/NC)_i$ using $AgNO_3$@DDAB and $AgNO_3$@DDA/DDAB, respectively. The blue curves shows the optical absorption spectrum of 3.4 InAs NCs doped with 500 $(Ag/NC)_i$ using $AgNO_3$@DDA.

Figure S12 depicts the XANES spectra, as collected from the In (a-b) and As (d-f) K-edges for all three NC sizes (3.4nm, 4.8nm, and 6.6mn, respectively) exhibits no change in the position of the absorption edge of both As and In edges when increasing the $(Ag/NC)_i$ ratios.

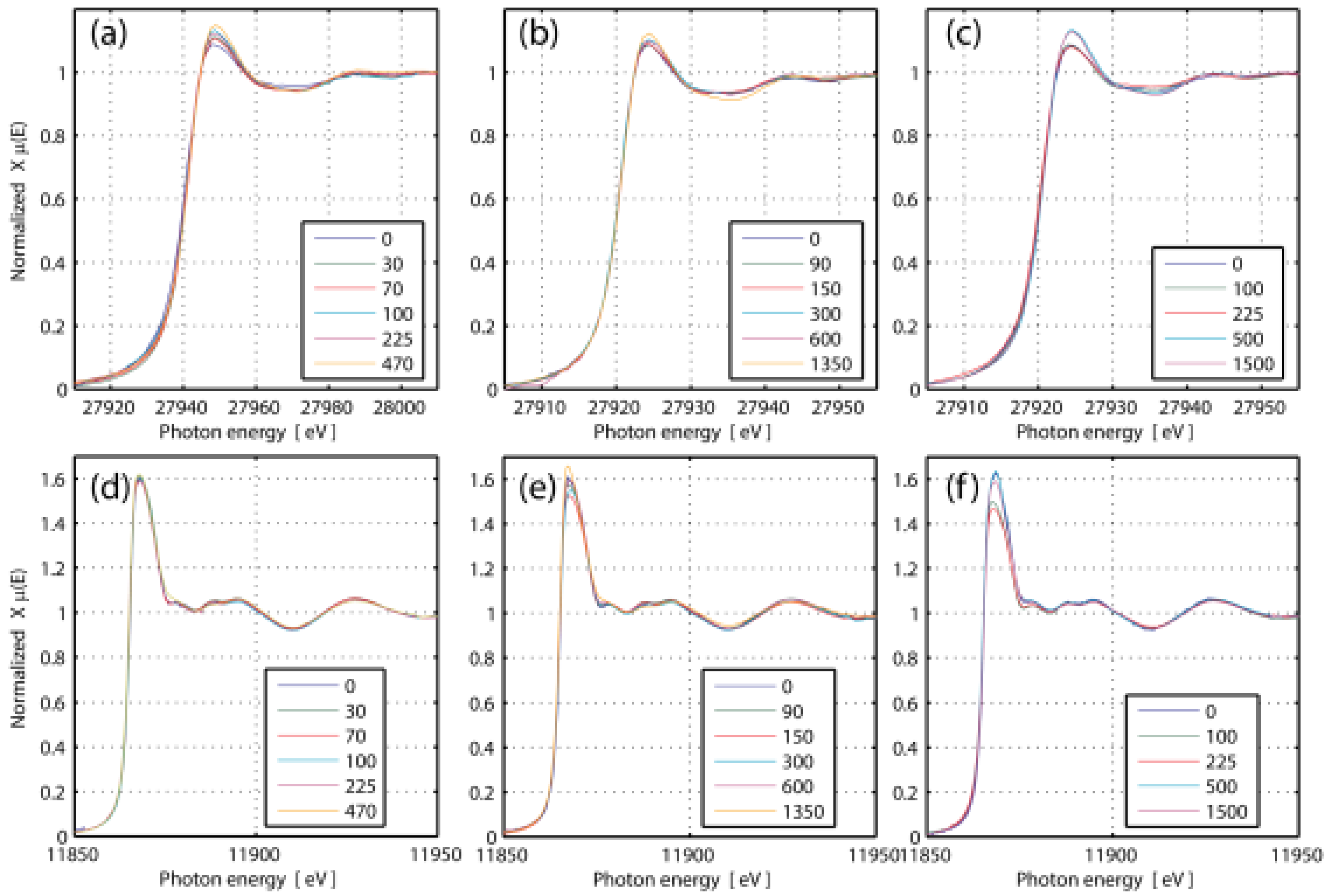

**Figure S12**. In (a-c) and As (d-f) K-edge XANES spectra for 3.4nm (left), 4.8nm (middle), and 6.6nm (right) Ag doped InAs NCs (legends depict the $(Ag/NC)_i$ ratios in the solution). The legend depicts the concentration of Ag atoms as calculated per NC in the reaction solution. All samples were precipitated and measured in the solid phase after spread on Kapton tape.

Figure S3 presents the K-space spectra of the EXAFS data as collected for the 4.8nm NC samples. The k-range used for the Fourier transform (Figure 6b and Figure S10b,c) was 2-13 Å$^{-1}$, 2-12 Å$^{-1}$, and 2-11.4 Å$^{-1}$ for the In, As, and Ag K-edges, respectively.

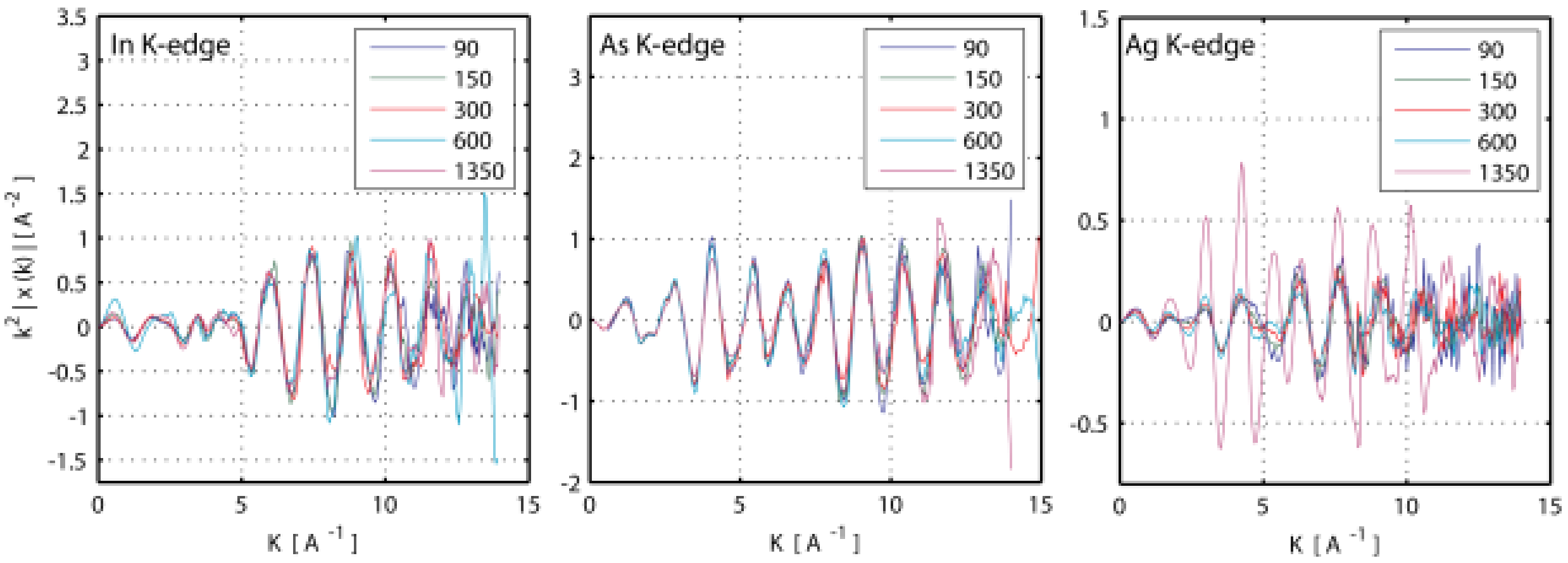

**Figure S13.** K-space spectra of the EXAFS data in In (left), As (middle) and Ag (right) K-edges as collected for 4.8nm InAs NCs doped with various concentrations of Ag. Legends denote the $(Ag/NC)_i$ ratios in the solution.

Figure S14 shows the Fourier-transform magnitudes of the EXAFS data spectra for the 3.4nm (left), 4.8nm (middle), and 6.6nm (right) Ag doped NCs measured from the As (top) and In (bottom) K-edges. Visual examination of the EXAFS data indicates that for both As and In edges, the position of the main R-space peak is maintained and no new peaks evolve upon doping.

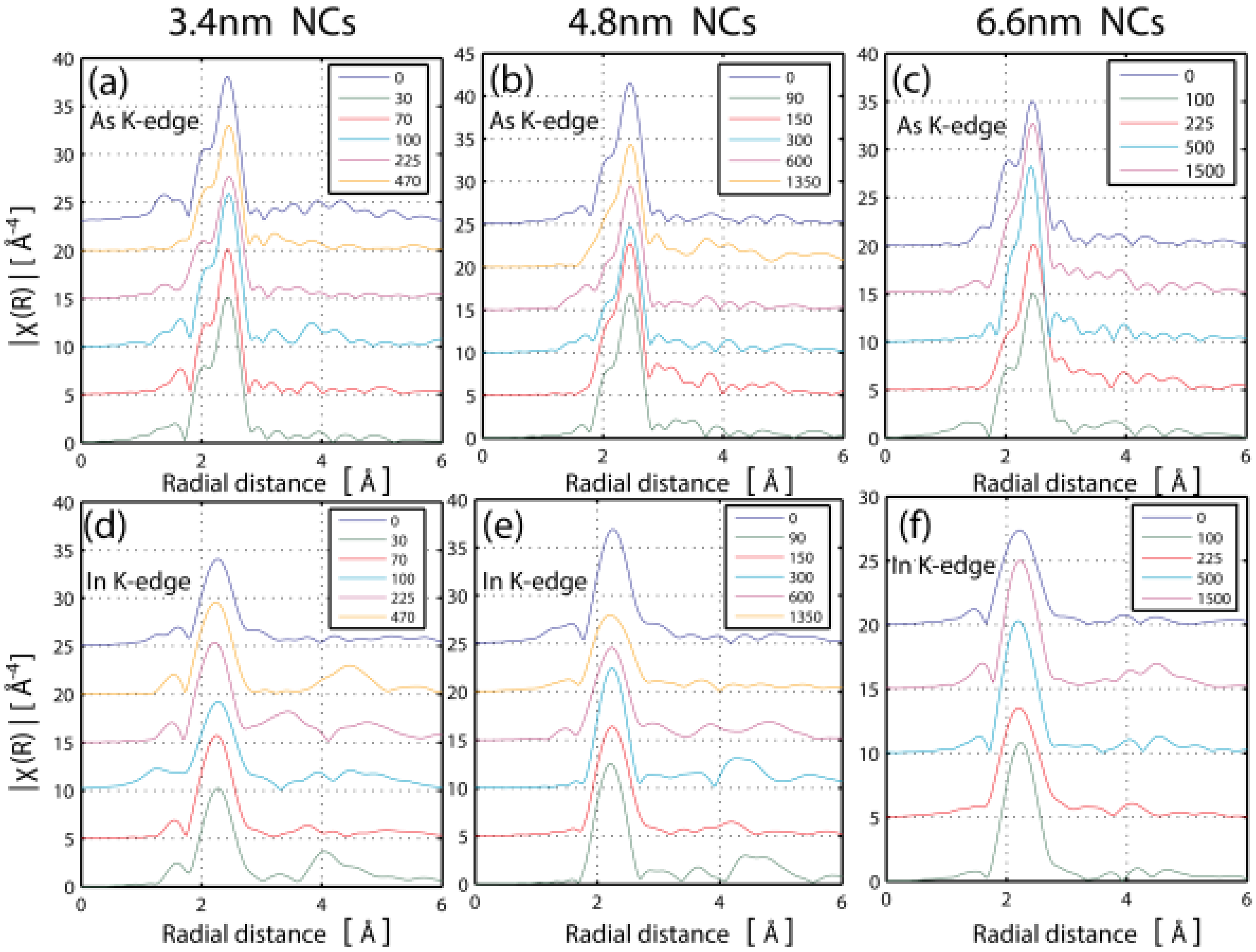

**Figure S14**. Fourier-transform magnitudes of the EXAFS spectra at different Ag concentrations in solution for As K-edge (a-c) and In K-edge (d-f) of the 3.4nm (left), 4.8nm (middle), and 6.6nm (right) Ag doped InAs NCs. The spectra are offset vertically for clarity of presentation. All legends depict the $(Ag/NC)_i$ ratios in the solution.

The structural model used to fit the EXAFS data of all Ag, In, and As K-edges is that of Ag being a substitutional impurity occupying an In lattice site (Fig. S15). The best-fit results of the EXAFS analysis for the different NC sizes doped with various $(Ag/NC)_i$ ratios are presented in Figure S16 and the obtained values of the adjustable parameters are listed in Table S2

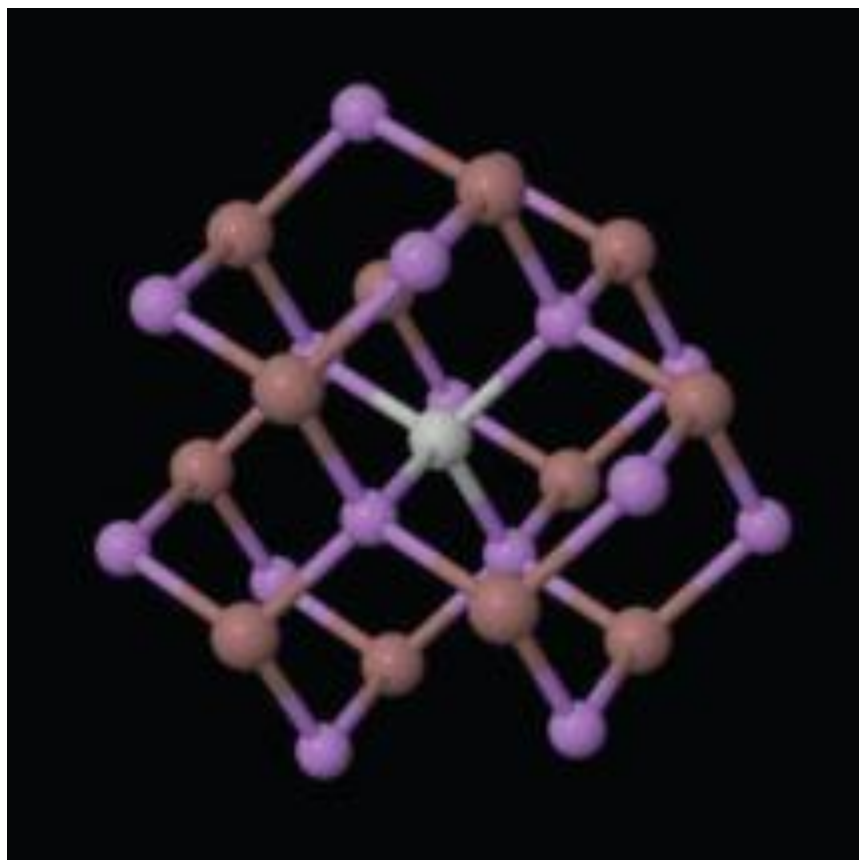

**Figure S15**. Illustrations of the substitutional site for a Ag (grey) impurity in an InAs (brown , purple, respectively) NC. The substitutional site is characterized by: N(As) = 4 and r(Ag−As) = 2.61 Å.

**Table S2**. Best-Fit Results for Structural Parameters of the Ag−As, Ag−Ag, and Ag-N Nearest Neighbor Pairs Obtained by Ag K-Edge Analysis for Various $(Ag/NC)_i$ ratios of 3.4nm, 4.8nm, and 6.6nm InAs NCs

| Sample | Dia. [nm] | bond | N | R(Å) | $\sigma^2$(Å$^2$) | $\Delta E_0$ |
|---|---|---|---|---|---|---|
| YA9F5-Ag30 | 3.4 | Ag-As | 2.1±0.6 | 2.523±0.018 | 0.009±0.002 | -5.7 ±4.3 |
| | | Ag-N | 1.9±0.8 | 2.826±0.097 | 0.009±0.002 | 7.6 ±4.1 |
| YA9F5-Ag70 | | Ag-As | 1.8±0.5 | 2.539±0.018 | 0.009±0.002 | -5.7 ±4.3 |
| | | Ag-N | 1.6±0.7 | 2.857±0.100 | 0.009±0.002 | 7.6 ±4.1 |
| YA9F5-Ag100 | | Ag-As | 1.6±0.6 | 2.541±0.020 | 0.009±0.002 | -5.7 ±4.3 |
| | | Ag-N | 0.9±0.9 | 2.845±0.138 | 0.009±0.002 | 7.6 ±4.1 |
| | | Ag-Ag | 1.4±0.7 | 2.861±0.038 | 0.016±0.002 | -0.8±0.8 |
| YA9F5-Ag225 | | Ag-As | 1.0±0.4 | 2.556±0.023 | 0.009±0.002 | -5.7 ±4.3 |
| | | Ag-N | 0.6±0.5 | 2.789±0.126 | 0.009±0.002 | 7.6 ±4.1 |
| | | Ag-Ag | 1.7±0.5 | 2.882±0.023 | 0.016±0.002 | -0.8±0.8 |
| YA9F5-Ag470 | | Ag-As | 0.7±0.2 | 2.605±0.027 | 0.009±0.002 | -5.7 ±4.3 |
| | | Ag-Ag | 4.2±0.6 | 2.850±0.012 | 0.016±0.002 | -0.8±0.8 |
| YA5F3-Ag90 | 4.8 | Ag-As | 1.7±0.4 | 2.531±0.014 | 0.008±0.002 | -5.8 ±3.3 |
| | | Ag-N | 1.2±0.7 | 2.838±0.103 | 0.008±0.002 | 8.5 ±4.3 |
| YA5F3-Ag150 | | Ag-As | 1.4±0.3 | 2.529±0.014 | 0.008±0.002 | -5.8 ±3.3 |
| | | Ag-N | 1.0±0.4 | 2.832±0.094 | 0.008±0.002 | 8.5 ±4.3 |
| | | Ag-Ag | 0.6±0.2 | 2.884±0.031 | 0.011±0.001 | -0.3±0.5 |
| YA5F3-Ag300 | | Ag-As | 1.2±0.4 | 2.549±0.018 | 0.008±0.002 | -5.8 ±3.3 |
| | | Ag-N | 0.8±0.7 | 2.872±0.129 | 0.008±0.002 | 8.5 ±4.3 |
| | | Ag-Ag | 0.9±0.5 | 2.868±0.037 | 0.011±0.001 | -0.3±0.5 |
| YA5F3-Ag600 | | Ag-As | 0.9±0.3 | 2.561±0.016 | 0.008±0.002 | -5.8 ±3.3 |
| | | Ag-N | 0.6±0.4 | 2.841±0.112 | 0.008±0.002 | 8.5 ±4.3 |
| | | Ag-Ag | 1.4±0.3 | 2.869±0.013 | 0.011±0.001 | -0.3±0.5 |
| YA5F3-Ag1350 | | Ag-As | 0.7±0.2 | 2.622±0.026 | 0.008±0.002 | -5.8 ±3.3 |
| | | Ag-Ag | 7.4±0.6 | 2.847±0.006 | 0.011±0.001 | -0.3±0.5 |
| YA8F2-Ag100 | 6.6 | Ag-As | 1.8±1.0 | 2.531±0.028 | 0.008±0.003 | -1.6 ±4.6 |
| | | Ag-N | 2.0±1.3 | 2.766±0.157 | 0.008±0.003 | 5.4 ±7.8 |
| YA8F2-Ag225 | | Ag-As | 2.1±1.1 | 2.546±0.029 | 0.008±0.003 | -1.6 ±4.6 |
| | | Ag-N | 2.2±1.2 | 2.781±0.159 | 0.008±0.003 | 5.4 ±7.8 |
| YA8F2-Ag500 | | Ag-As | 1.0±0.7 | 2.563±0.030 | 0.008±0.003 | -1.6 ±4.6 |
| | | Ag-N | 0.7±0.8 | 2.745±0.225 | 0.008±0.003 | 5.4 ±7.8 |
| | | Ag-Ag | 1.6±0.8 | 2.894±0.045 | 0.016±0.002 | -0.7±0.8 |
| YA8F2-Ag1500 | | Ag-As | 1.3±0.5 | 2.616±0.027 | 0.008±0.003 | -1.6 ±4.6 |
| | | Ag-Ag | 5.0±0.8 | 2.845±0.017 | 0.016±0.002 | -0.7±0.8 |

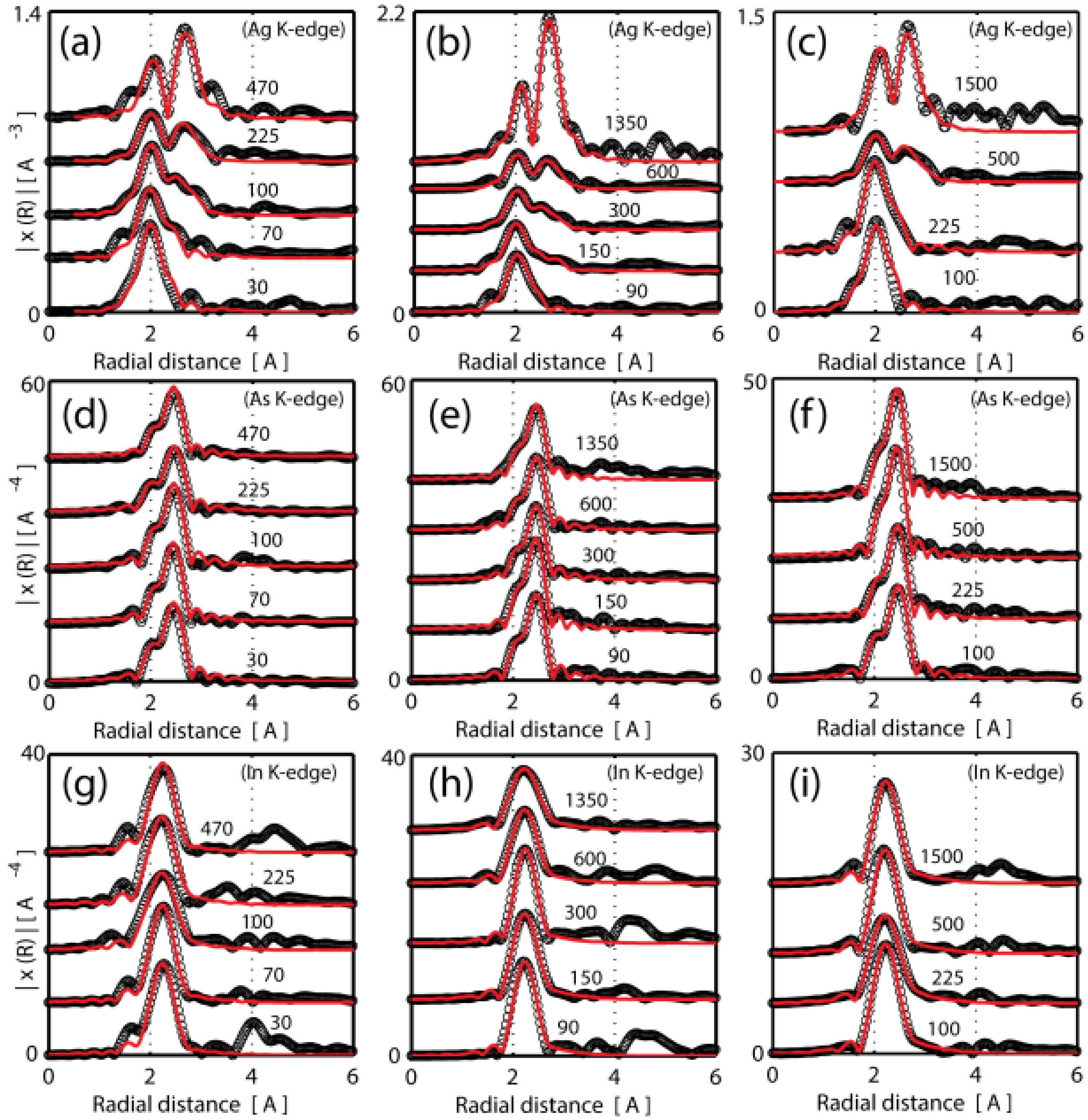

**Figure S16**. Fourier-transform magnitudes of the EXAFS spectra (open symbols) and the best fit results (red curve) for the Ag K-edge (a-c), As K-edge (d-f), and the In K-edge (h-j) of the 3.4nm (left), 4.8nm (middle), and 6.6nm (right) Ag doped InAs NCs. Each plot is accompanied with the corresponding $(Ag/NC)_i$ ratio that was used in the solution.

**Table S3** Reduced Chi-square and R-factor obtained by using models with (Model 1) and without (Model 2) including Ag-N contributions. For the model (Model 2) without including Ag-N contribution, the coordination number of Ag-As was constrained to be 4.

| Dia. [nm] | Model | Reduced Chi-square | R-factor |
|---|---|---|---|
| 3.4 nm | 1 | 32.59 | 0.016 |
| | 2 | 50.14 | 0.071 |
| 4.8nm | 1 | 26.45 | 0.021 |
| | 2 | 81.16 | 0.083 |
| 6.6nm | 1 | 24.85 | 0.031 |
| | 2 | 61.24 | 0.052 |

Figure S17 presents the inter-atomic (R) distance of the Ag-As bond (green) and In-As bond (black) as a function of the $(Ag/NC)_a$ ratio for the 3.4nm (left) and 6.6nm (right) doped InAs NCs. Similar to the 4.8nm InAs sample, the In-As bond length is unaffected by the doping whereas the Ag-As bond length slightly decreases when increasing the Ag concentration.

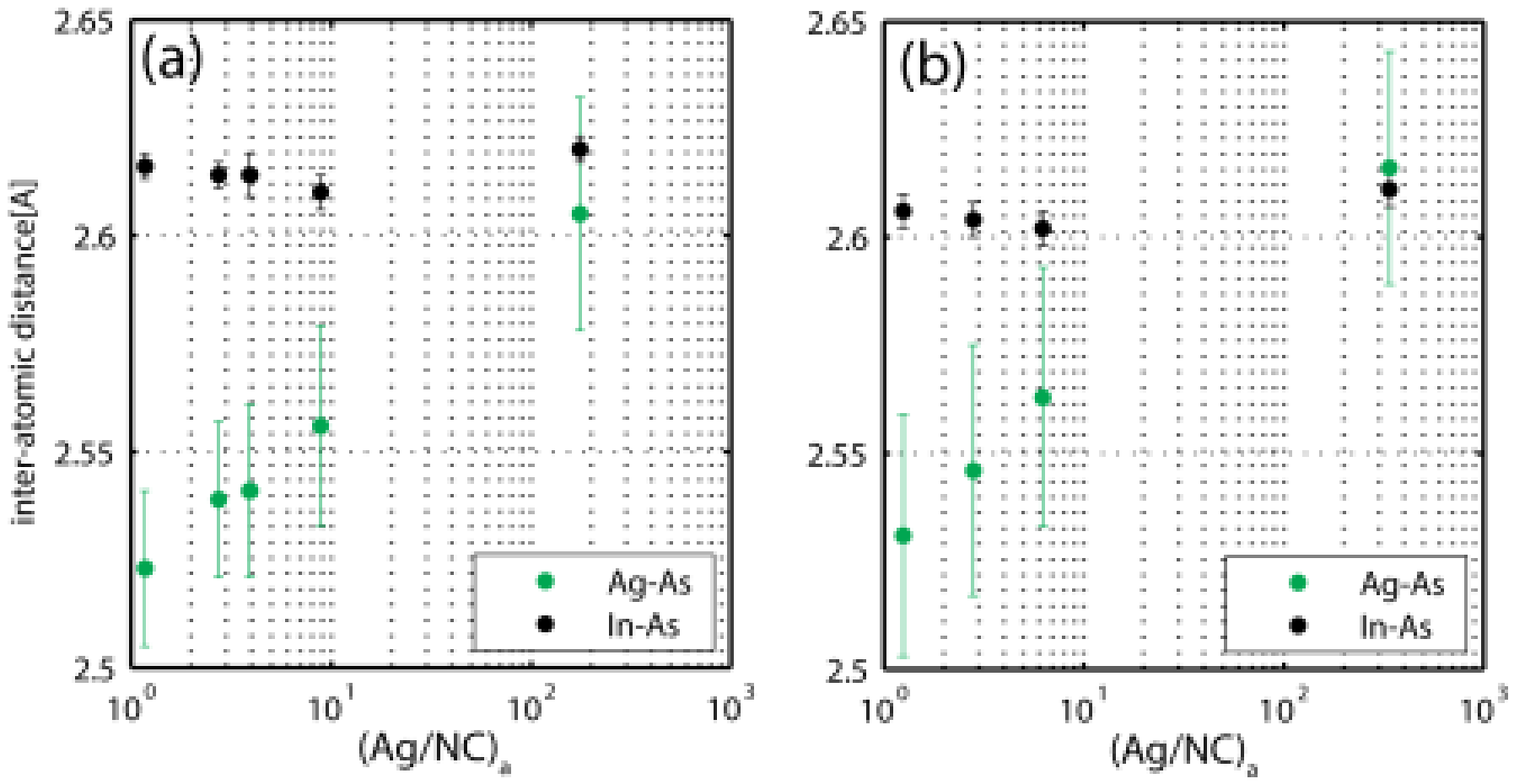

**Figure S17**. Ag-As (green) and In-As (black) inter-atoms distances as a function of the $(Ag/NC)_a$ ratio (obtained by ICP-MS). (left) for the 3.4nm doped InAs NCs and (right) for the 6.6nm doped InAs NCs.

Figure S18 shows how the average coordination number evolves for an ideal, truncated-cuboctahedral, Ag cluster as a function of the cluster diameter (a) and the total number of atoms in the cluster (b). The values were calculated using the following equations:

(1) $$N_{atoms} = \frac{5}{3}L^3 + 4L^2 + \frac{10}{3}L + 1$$

(2) $$N_{bonds} = \frac{20L^3+21L^2+7L}{2}$$

(3) $$CN = 2\frac{N_{bonds,TCO}}{N_{atoms,TCO}}$$

(4) $$d = 2Lr_{Ag-Ag}$$

Where $N_{atoms}$, $N_{bonds}$, CN, and d are: the number of atoms, number of atomic bonds, the average coordination number, and the diameter of the Ag cluster, respectively. L is defined as the order of the cluster representing the number of inter-atomic distances along the cluster edge, and $r_{Ag\text{-}Ag}$ is the bulk bond length for metallic Ag (2.87Å). Best fit results obtained by EXAFS analysis (Table S1) for the highest doping concentrations for the 3.4nm, 4.8nm, and 6.6nm samples yield an average Ag-Ag coordination numbers of 4.2±0.6, 7.4±0.6, and 5.0±0.8, respectively. From figure S18 we learn that if these were ideal Ag clusters they would have a characteristic size of 5Å and consist of just a few atoms.

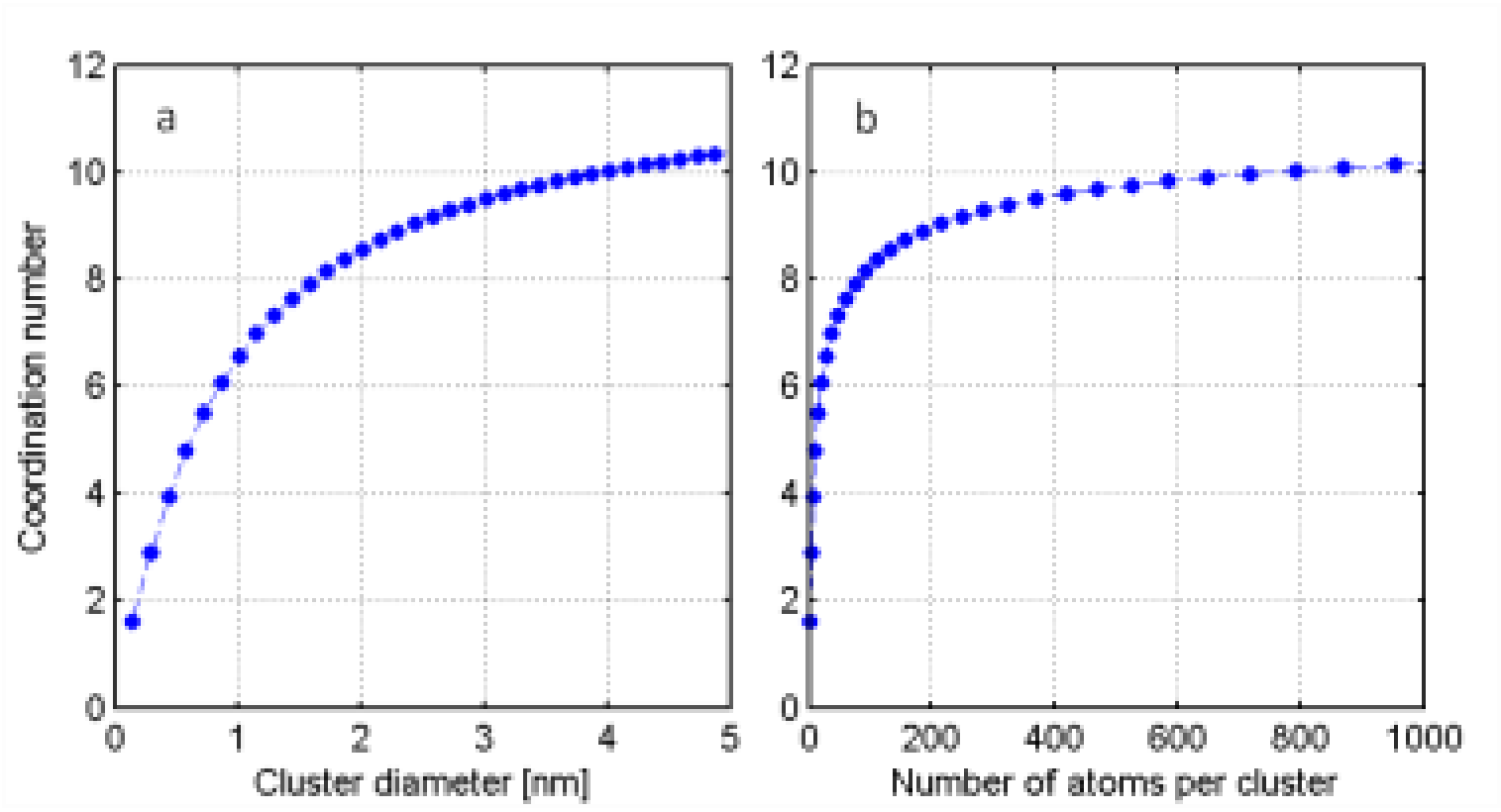

**Figure S18.** Theoretical calculation of (a) the average coordination number as a function of the cluster diameter, and (b) the average coordination number as a function of the number of atoms in the cluster, for an ideal truncated-cuboctahedral Ag cluster.

Figure S19 shows the XRD data of 3nm (a) and 7nm (b) InAs NCs at different $(Ag/NC)_i$ ratios. It can be seen that for all $(Ag/NC)_i$ ratios the main InAs reflections are preserved, suggesting that the addition of Ag impurities induces relatively small disorder in the NC lattice. Furthermore, the lack of new peaks in the spectrum for the low $(Ag/NC)_i$ ratios indicates that no metallic Ag is formed. However, the evolution of a new diffraction peak may be observed at 2θ=15 deg. for higher Ag levels, identified as the (111) Ag reflection.

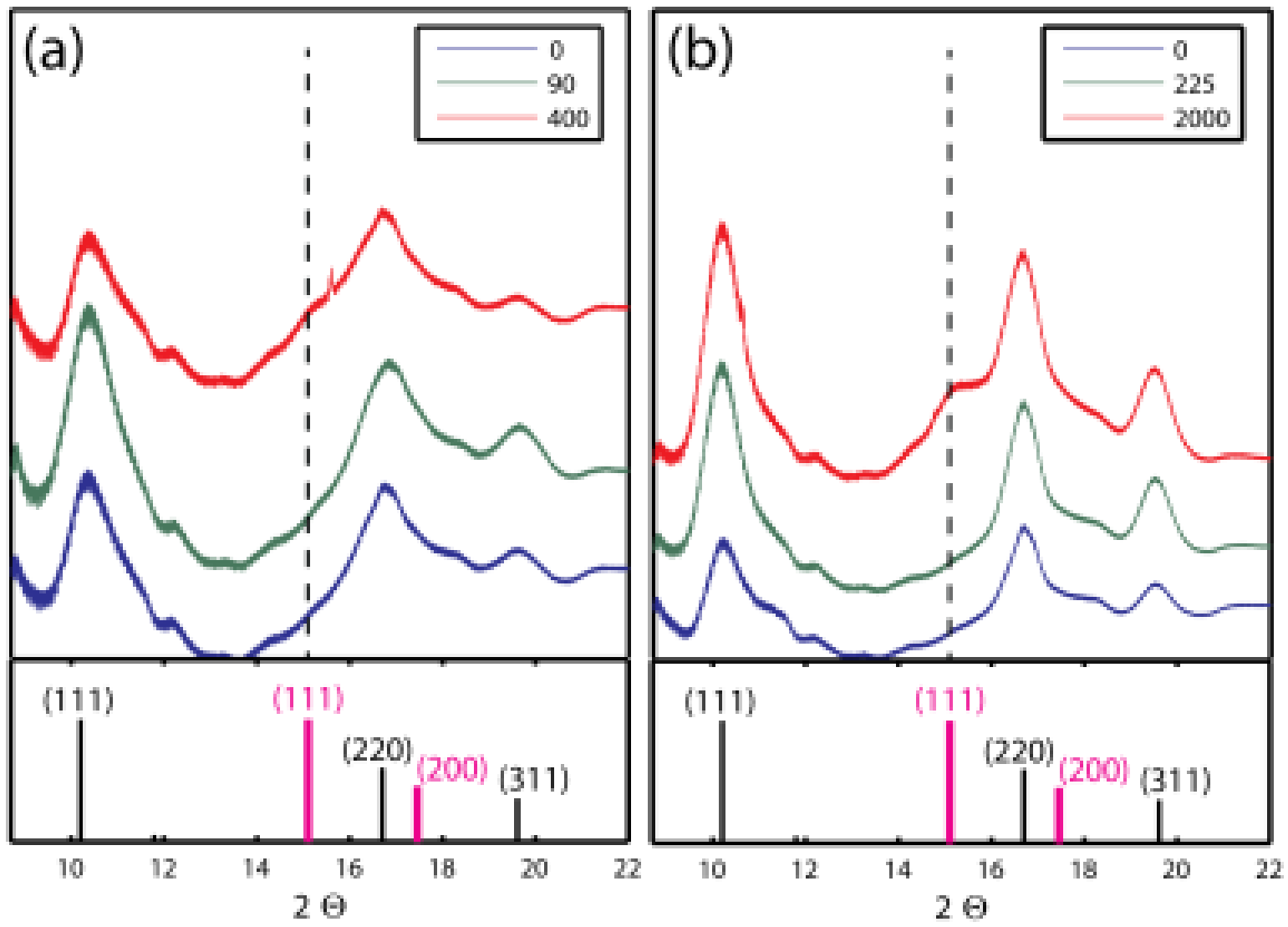

**Figure S19**. Synchrotron based powder XRD measurements (λ=0.62Å) for (a) 3.4nm and (b) 6.6nm InAs NCs doped to various $(Ag/NC)_i$ ratios in solution. Bottom panes represent the standard diffraction peaks for InAs zincblende lattice (black bars) and Metallic Ag FCC (magenta bars). The dashed lines represents the position of the (111) Ag crystal plane.

---

[1] Tauc, J. In Amorphous and Liquid Semiconductors; Plenum: New York, 1974.

[2] Mocatta, D.; Cohen, G.; Schattner, J.; Millo, O.; Rabani, E.; Banin, U. Heavily Doped Semiconductor Nanocrystal Quantum Dots. *Science* **2011**, 332, 77−81.